\documentclass[12pt,a4paper]{article}

\usepackage{epsfig}

\def\be{\begin{equation}}
\def\ee{\end{equation}}
\def\bdm{\begin{displaymath}}
\def\edm{\end{displaymath}}
\def\bea{\begin{eqnarray}}
\def\eea{\end{eqnarray}}
\def\ba{\begin{array}}
\def\ea{\end{array}}

\def\arsinh{\mbox{arsinh}}
\def\arcosh{\mbox{arcosh}}
\newcommand{\RR}{R}

\newcommand{\avp}[1]{\langle #1 \rangle}
\newcommand{\av}[1]{\avp{#1}_{\Lambda}}
\def\llangle{\langle\!\langle}
\def\rrangle{\rangle\!\rangle}

\newcommand{\Bild}[4]{
\begin{figure}[htb]
  \begin{center}
    \leavevmode
    \epsfig{file=bilder/#2,height=#1cm}
    \caption{{\small #3}}
    \label{#4}
  \end{center}
\end{figure}}

\begin{document}

\title{\bf Random Tilings: \\ Concepts and  Examples}
\author{{\sc Christoph Richard}, {\sc Moritz H\"offe},\\
 {\sc Joachim Hermisson} and {\sc Michael Baake} \\[2mm] 
{\small Institut f\"ur Theoretische Physik, Universit\"at T\"ubingen,} \\
{\small Auf der Morgenstelle 14, D-72076 T\"ubingen, Germany}}

\maketitle

\begin{abstract}
We introduce a concept for random tilings which, comprising the conventional one,
is also applicable to tiling ensembles without height representation.
In particular, we focus on the random tiling entropy as a function of the tile densities.
In this context, and under rather mild assumptions, we prove a generalization of
the first random tiling hypothesis which connects the maximum of the entropy 
with the symmetry of the ensemble.
Explicit examples are obtained through the re-interpretation of several exactly solvable
models.
This also leads to a counterexample to the analogue of the second random tiling
hypothesis about the form of the entropy function near its maximum.
\end{abstract}

\subsection{Introduction}

Perfect quasiperiodic tilings provide useful tools for the description of certain
thermodynamically stable quasicrystals \cite{GKB,JA,NB}, at least for their
averaged structure.
However, it is widely accepted that the perfect structure is an idealization, and that
a systematic treatment of defectiveness is necessary for all but a few quasicrystals
\cite{BA,E,Hen91}.
In particular, starting from the perfect structure, the inclusion of local defects
\cite{BA} is helpful to overcome a number of pertinent problems, ranging from the
observed degree of imperfections \cite{BO} to the impossibility of strictly local
growth rules for perfect quasiperiodic patterns \cite{O}.

An alternative approach relies on a stochastic picture and adds an entropic side to
the stabilization mechanism. 
Its most prominent representatives are usually formulated in terms of so-called
random tilings \cite{E,Hen88,Hen91}.
It has long been argued that such stochastic tilings should provide a more realistic
description both of the structures with their intrinsic imperfections and of their
stabilization, which is then mainly entropic in nature.

Recently, improved methods have been suggested \cite{JB} to test this hypothesis in
practice, and the results support the claim that such models are realistic and
relevant \cite{JBR}.
Further impetus is given by the very recent progress to grow (almost) equilibrium
random tilings on the computer \cite{JE} in suitable grand-canonical ensembles.

The concept of entropic stabilization applies to a wide class of --
not necessarily quasicrystalline --
tiling ensembles (defined through a set of proto-tiles and packing rules). 
However, within the original approach to random tilings (see \cite{Hen91} for
a review), only a (while interesting) rather special subset of these can be
described. This is because two concepts are mixed here that, in principle,
have nothing to do with each other: 
the mechanism of entropic stabilization and the concept of the
phason strain, the latter being used for the description of quasicrystalline order for
those tilings that allow a so-called height
representation. Hereby, the phasonic language is used throughout,
restricting the range of applicability.  

In this paper, we propose a more elementary and, at the same
time, more rigorous point of view which directly employs the densities of the 
tiles to describe the phase diagram. While allowing for
adequate generality, the new picture conforms to the other one,
if this exists -- at least, if the relation between the tile densities
and the strain parameters is locally linear.

This reformulation is, however, not done for mere aesthetic reasons, 
but has a number of advantages and leads to consequences of direct interest.
In particular, the generality of the approach now makes classes
of exactly solvable models accessible -- e.~g. examples compatible with 
{\em crystallographic} symmetries -- which already show a variety of 
interesting and typical phenomena.

For an explicit result, let us mention the two so-called random tiling hypotheses.
The validity of these hypotheses is the usual set-up for an elastic theory
in order to show the existence of a diffraction pattern with Bragg peaks for 
three-dimensional tilings, see \cite{Hen91} and, for a recent
discussion of experimental results, \cite{Cod97}.
The first hypothesis assumes that the point of maximum entropy in the phase space is one of
maximal symmetry. 
We will show below how it can be derived from more basic assumptions as a consequence,
and need not be stated as hypothesis.
The second hypothesis uses the phasonic picture of quasicrystals to establish a kind
of elastic theory. 
The hypothesis would now state that the entropy is locally a quadratic function of the
densities near its maximal value, and the Hessian can be interpreted as an 
entropic elasticity tensor.
However, this picture should to be taken with a pinch of salt:
we shall show a counterexample where the maximum is unique, but coincides with a second
order phase transition and is thus {\em not} of quadratic nature.

In the conventional setup much information, such as elastic constants and diffraction
properties, are extracted from the correlations of the height variables.
The fundamental variable analogous to the height correlations in our picture turns out 
to be the {\em covariance matrix} of the tile numbers.

Let us sketch how the paper is organized.
The following section is devoted to the introduction of our concept and the thermodynamic 
formalism used.
After a brief discussion of the class of (generalized) polyomino tilings \cite{Gol94}, 
which comprises all our examples,
we will illustrate our concepts by simple examples in one dimension.
In the section on symmetry versus entropy, we present an argument on how to
derive the first random tiling hypothesis from elementary assumptions.
This is then substantiated by various planar examples, where our focus is on the
re-interpretation of exactly solvable models which thereby gain a slightly
different interpretation and, eventually, even another application. 
Our examples are taken from dimer models, but also the three-colouring model on the 
square lattice and hard hexagons are discussed.

\subsection{Setup and preliminaries}

In this section, we describe the way how we deal with random tilings. 
Since we do not want to use the special and somewhat restrictive phasonic
picture for our fundamental definitions, these will differ from those given elsewhere
\cite{Hen91}.

As a {\em (random) tiling} we define a face-to-face space filling with tiles from a
finite set of prototiles, without any gaps or overlaps. 
There might be a number of additional local packing rules specifying the 
allowed tilings. 
The number of allowed patches should increase exponentially with the volume of the 
patch (and hence with the number of tiles), resulting in a positive entropy density 
for the tiling ensemble, which we then call {\em random tiling ensemble}. 
For the tiling ensemble being homogeneous, prototile shapes and packing rules
should admit tilings containing all sorts of tiles (with positive density),
otherwise the ensemble splits into different subensembles that have to be taken
into account separately.

With this setting, random tilings in the sense of \cite{Hen91} are included 
as a proper subset.
In order to give the thermodynamic limit a precise meaning, we formalize
the definition as follows.

For a random tiling of the space $\RR^d$, its {\em prototiles} ${\cal{T}}_i$ are 
compact, connected subsets of $\RR^d$ of positive volume $l_i$, homeomorphic to balls.
Let $\Lambda \subset \RR^d$ be given as connected, compact set of positive 
volume $V(\Lambda)$.
We now have to explain our notion of a patch.
If $P_\Lambda$ is a collection of translated prototiles,
we call $P_\Lambda$ a {\em $\Lambda$-patch} iff $P_\Lambda$ is a covering 
of $\Lambda$ (without gaps) such that each element of $P_\Lambda$ has a nonempty 
intersection with $\Lambda$, and every two elements have empty common interior.
We call two $\Lambda$-patches {\em equivalent} iff they are 
translates of each other.
For fixed prototile numbers $n_1,\ldots,n_M$ (with corresponding
densities $\rho_i = l_i n_i/V(\Lambda)$), let us denote the number of non-equivalent
$\Lambda$-patches by $g_\Lambda (n_1,\ldots,n_M)$.

It should be stressed here that, in principle, we can impose any type of boundary 
conditions for our definitions.
However, {\em free boundary conditions} are the natural ones to choose from a 
physical point of view -- a different choice may indeed lead to deviating
results; fixed boundary conditions in particular seem to be too restrictive
in most cases \cite{Pro97,E1,LR91}.
On the other hand, most of the exact results have been obtained by use
of periodic boundary conditions, and it has to be shown, for each example,
that the results also apply to the case of free boundary conditions.

We will use a grand-canonical formulation in the following, in difference
to the (usual) canonical treatment of tiling ensembles \cite{Hen91,LPW92}, as we
believe that an ensemble with fluctuating tile numbers is the adequate setup 
to cover processes such as local rearrangements or tiling growth.
Moreover, this formulation will prove advantageous for our arguments and practical
calculations.

As we are interested in configurational entropy only, we will assign equal (zero)
energy to every prototile.
In this situation, the chemical potentials of the prototiles are the only 
parameters remaining.
As variables conjugated to the densities, we introduce
chemical potentials $\mu_i$ as well as activities $z_i = e^{\mu_i}$ for 
every prototile,
\footnote{
We suppress extra prefactors $\beta$ in all definitions below.
}
($i=1,\ldots,M$), and define the grand-canonical partition function to be the
configuration generating function
\be 
{\cal Z}_\Lambda(\mu_1,\ldots,\mu_M)=\sum_{n_1,\cdots,n_M} 
g_\Lambda(n^{}_1,\ldots,n^{}_M) z_1^{\ell^{}_1 n^{}_1}\cdots z_M^{\ell^{}_M
n^{}_M}. \label{form:gf}
\ee

Since we are interested in the entropic behaviour of large systems, 
we define the grand-canonical potential
\footnote{
If Boltzmann-factors are taken instead of activities, this corresponds to the
(canonical) free energy in statistical mechanics.
}
per unit volume to be the limit 
\be 
\phi(\mu_1,\ldots,\mu_M) = \lim_{\Lambda\to\infty} \frac{1}{V(\Lambda)}\log
{\cal Z}_\Lambda(\mu_1,\ldots,\mu_M). \label{form:grand}
\ee
Here, $\{\Lambda\}$ is any collection of connected, compact subsets of $\RR^d$ of 
positive volume $V(\Lambda)$ such that $ \lim V(\Lambda) = \infty$ and 
$\lim B(\Lambda)/V(\Lambda) = 0$, where $B(\Lambda)$ denotes (the maximum) over the
{\em volume of the boundary tiles} of any $\Lambda$-patch, in the same spirit as
the van Hove limit taken for ensembles in statistical mechanics \cite{R}.
For models of physical relevance the thermodynamic limit defined that way should exist
for all physical quantities and be independent of the special choice of the limit 
sequence $\{\Lambda\}$.

The mean (volume) densities of the different prototiles in this ensemble can be 
computed as
\be 
\bar{\rho}_i(\mu_1,\ldots,\mu_M) = \lim_{\Lambda \to \infty} 
\frac{1}{V(\Lambda)}\av{l_i n_i} = 
\frac{\partial \phi(\mu_1,\ldots,\mu_M)}{\partial \mu_i}
\quad (i=1,\ldots,M), \label{form:Dichte} 
\ee
where $\av{\mbox{ }}$ denotes the finite-size ensemble average for 
given chemical potentials $\mu_1,\ldots,\mu_M$. 

Note that all physical quantities defined here are {\it ensemble
averages}.
In cases where these quantities differ from corresponding
values of a {\it typical} representative of the ensemble,
one has to reflect whether the grand-canonical setup fits
the experimental situation, or if a canonical ensemble is the
correct choice.
Bear in mind, however, that the van Hove limit is not properly defined
for single tilings that are not self-averaging.
Thus either canonical and grand-canonical thermodynamics 
coincide, or the the canonical picture is likely not to be
well defined at all.

So far, the chemical potentials and also the conjugated (mean) 
densities do not form an {\em independent} set of macroscopical parameters 
to describe the ensemble. 
We always have $\sum_{i=1}^M \bar{\rho_i}=1$ as normalization constraint, 
and in general, the tile geometries and packing rules may eventually impose further
constraints on the densities. 

In order to make the notion of macroscopical independence precise,  let 
$v = \sum v_j n_j \neq 0$ be an arbitrary linear combination of some 
prototile numbers $n_j$, with real coefficients $v_j$. 
We call the accompanying set of tile densities $\{\bar{\rho_j}\}$ 
{\em independent} in the thermodynamic limit, iff the variance per
unit volume of $v$ is strictly positive for any choice of the $v_j$:
\be
\llangle v \rrangle := \lim_{\Lambda\to \infty} \frac{1}{V(\Lambda)} 
\av{(v -\av{v})^2} \;\; >\;  0 \label{form:variance}.
\ee
The {\em degrees of freedom} of the tiling ensemble are given by a maximal 
set of independent densities.
One checks that, with this definition, normalization reduces the independent densities by
one, and there is no additional freedom due to surface effects or
subclasses of the ensemble that do not contribute entropically.

In general, since the average taken in (\ref{form:variance}) depends on the
chemical potentials, the variance of some $v$ may vanish only for a special choice of
the potentials leading to a reduction of independent densities at isolated points
of the phase space.
However, no such thing will happen in any of the examples we discuss in
this sequel and, for simplicity of the presentation, we exclude this possibility
in the following.

Let us mention a consequence of this definition for the covariance matrix of the tile 
numbers (which corresponds to the isothermal compressibility in the thermodynamic situation),
in the thermodynamic limit:
\be \label{kappa}
\kappa_{ij} = \lim_{\Lambda \to\infty} 
\frac{1}{V(\Lambda)} \bigg( \av{l_i n_i l_j n_j} - \av{l_i n_i}\av{l_j n_j} \bigg)
= \frac{\partial\bar{\rho_i}}{\partial \mu_j}
\qquad (i,j=1,\ldots,M). \label{form:covariance}
\ee
This matrix describes tile fluctuations around the mean density and correlations among
different tiles. 
A straightforward calculation shows that (\ref{kappa}) is indeed strictly positive 
iff it is restricted to the subspace of density parameters that had been chosen 
independently according to the above definition.

In general, a linear constraint on the tile numbers of the form
\be 
\sum_{j=1}^M v_j (l_j n_j) = c V(\Lambda) + {\cal{O}}( V^\alpha(\Lambda) ) 
\qquad (\alpha < 1, \; c \in \RR)
\ee
leads to linear variation of the grand-canonical potential in certain coordinate directions: 
\be 
\phi( \mu_1 + v_1 t, \ldots, \mu_M+v_M t) = \phi(\mu_1,\ldots,\mu_M) + c t \qquad (t \in \RR).
\label{form:cons} 
\ee
For each constraint of this form, we can shift a single chemical potential to zero
in the grand-canonical potential by a suitable choice of $t$.

Let us choose an independent subset of macroscopic parameters, 
$\bar{\rho}_1,\ldots,\bar{\rho}_k$ say, which we abbreviate as $\{\bar{\rho}\}$,
and write the random tiling entropy as a function of these independent densities.
According to the considerations above, this can be done as follows.
We set $\mu_{k+1}=\cdots=\mu_M=0$ in  (\ref{form:gf}), i.e. we normalize the grand-canonical
potential by an appropriate choice of the origin of the energy scale.
We call the normalized grand-canonical potential $\phi$ again.
From (\ref{form:cons}) and (\ref{form:Dichte}) it is readily verified that
the normalization of $\phi$ leaves the independent densities invariant.

The (grand-canonical) entropy per unit volume as a function of the independent (mean)
tile densities is obtained as Legendre transform of the (normalized) 
grand-canonical potential 
\footnote{
Under the above assumption, this definition coincides with the micro-canonical
one \cite{P}; 
for a number of examples in the sequel this can be shown explicitly.}
with respect to the conjugated chemical
potentials,
\be
s(\bar{\rho_1},\ldots,\bar{\rho_k}) = \phi(\mu_1(\{\bar{\rho}\}),\ldots,\mu_k(\{\bar{\rho}\}))
 - \sum_{i=1}^{k} \bar{\rho_i} \mu_i(\{\bar{\rho}\}). \label{form:entropie}
\ee
Note that the Legendre transform is well defined owing to the non-singularity 
of the matrix
\be \frac{\partial^2 \phi(\mu_1,\ldots,\mu_k)}{\partial \mu_i\partial \mu_j} 
= \kappa_{ij} \qquad (i,j=1,\ldots,k). \ee

Let us locate the maximum of the entropy curve as a function of the independent
densities $\bar{\rho}_1,\ldots,\bar{\rho}_k$. 
We have by definition
\be
\frac{\partial s(\{\bar{\rho}\})}{\partial \bar{\rho}_j} = - \mu_j(\{\bar{\rho}\})
  \quad (j=1,\ldots,k).
\ee
It follows that the entropy is extremal, $s_{\rm max} = \phi$, at $\mu_1=\cdots=\mu_k=0$. 
(For the grand-canonical potential (\ref{form:grand}) this point occurs at
$\mu_1 = \cdots = \mu_M = 0$.)
This is the situation of the maximal random ensemble where each configuration 
is weighted equally.
Let us take a closer look at the Hessian, the matrix of the second derivative
\be 
\frac{\partial^2 s(\{\bar{\rho}\})}{\partial \bar{\rho}_i\partial \bar{\rho}_j} =
- \frac{\partial \mu_j(\{\bar{\rho}\})}{\partial\bar{\rho}_i} =
-\left(\kappa^{-1}\right)_{ij} \qquad (i,j=1,\ldots,k). 
\ee
If $s$ is locally $C^3$, we conclude that the entropy is a concave function everywhere.
On the other hand, also in the vector space of the independent
densities, the variance of $v$ and hence the covariance matrix
(\ref{kappa}) may diverge in the thermodynamic limit for certain
values of the chemical potentials. 
This generally occurs at phase transitions.
In this case the Hessian of the entropy has 0 as an eigenvalue in the 
large system limit; 
we will give an example below where this happens at the point of maximum entropy. 
For all other situations, the shape of the entropy curve near its maximum is completely 
determined by the Hessian.

We interpret the negative of the Hessian, at the point of maximum entropy,
as tensor of {\em entropic elasticity}.
Without any further symmetry, we refer to its eigenvalues as the {\em elastic constants}
of the random tiling ensemble.
We will define a symmetry-adapted version in the section on symmetry.

\subsection{Polyomino tiling ensembles}

A class of (crystallographic) tilings that allow a re-interpretation of many exactly
solved models are the (generalized) polyomino tilings, see  \cite{Gol94, LR91} for
definitions and background material.
Their prototiles are taken from one or combinations of several elementary cells
of a given periodic graph.
There may be as well certain packing rules, resulting in restricted random
tiling ensembles.
We will give a number of examples of tilings consisting of monominoes and dominoes.

The simplest class of random tilings are tilings of (coloured) monominoes
without further restrictions.
For lattices of any dimension, this results in Bernoulli ensembles \cite{Pet83}. 
A natural way to impose packing rules for these models is to exclude the occurrence
of clusters with moniminoes of the same colour.
For nearest-neighbour exclusion on lattices, the restriction is hierarchical:
the tiling can be regarded as being built from consecutive layers of ``hypertilings''
with additional constraints along the layer direction.
In this case, the entropy is bounded by the entropy of the corresponding hypertiling
ensemble.
Several two-dimensional examples are discussed below.

We should mention the one-to-one correspondence between 
polyomino tilings on a periodic graph $\Lambda$ and polymer arrangements 
on the dual cell complex $\Lambda^*$ (the so-called Delone complex).
This is illustrated in figure \ref{fig:dual} for a simple case, the monomer-dimer
system on the square lattice.
\Bild{4}{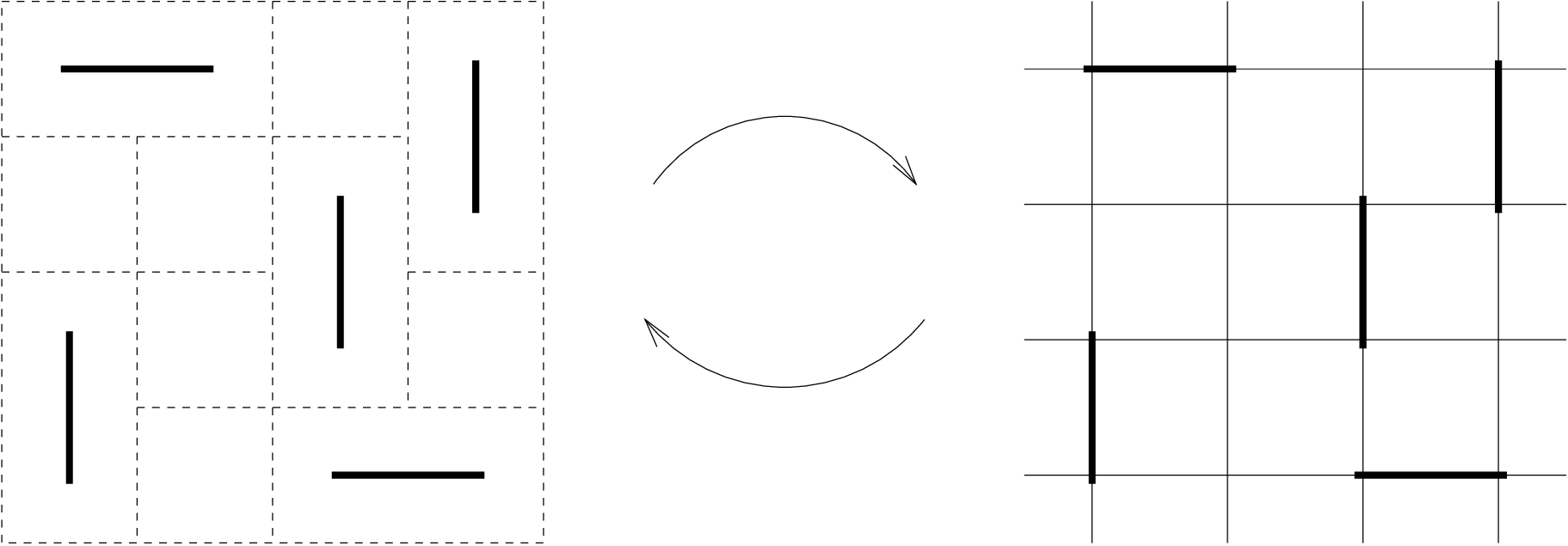}{Each polyomino tiling is a polymer configuration on the dual 
cell complex.}{fig:dual}

Owing to this connection, the (pure) domino problem can be solved in a number of cases
in two dimensions, as the dual fully packed dimer problem allows the exact solution 
on planar graphs by means of Pfaffians 
\footnote{
For an introduction into the Pfaffian method, we refer to Appendix E of \cite{T}.
}.
This has been shown by Kasteleyn in \cite{K,K2}, where he computed the entropy of 
dimers on the square lattice.
It is even possible to compute densities of finite patches explicitely, which has 
recently been shown by Kenyon \cite{Ken97}.
We will treat two examples representing the different types of critical behaviour
of the solvable domino systems \cite{N}:
systems of Onsager type (with logarithmic singularity of the covariance matrix) and
systems of Kasteleyn type (with square-root singularity of the covariance matrix).

Let us consider monomino-domino tilings without restriction next.
The one-dimensional system will be discussed below.
In two dimensions, there are various connections with other models,
e.~g. for the triangle-rhombus tiling on the triangular lattice \cite{Wu74},
but, to our knowledge, no explicit exact solutions.
On the other hand, as follows from the theory of monomer-dimer systems \cite{HL},
phase transitions can only occur for tilings with {\em vanishing} monomino density,
regardless of the dimensionality of the system -- these phases are the (pure) domino 
tilings mentioned above.
There is at least one monomino-domino tiling with constraints that allows for
exact solution in two dimensions:
the tiling on the hexagonal lattice together with exclusion of neighbouring 
dominoes on different sublattices.
This tiling ensemble is equivalent to the ensemble of hard hexagons \cite{B2,B},
see also below.

\subsection{Unrestricted and one-dimensional examples}

The simplest class of random tilings, built from $M$ coloured monominoes on
an arbitrary lattice, can be described purely combinatorially. 
There are no local constraints or packing rules, and thus no global constraints 
for the densities besides normalization. 
One application of this model is the description of unrestricted
chemical disorder in multicomponent alloys \cite{Bal91}.

The thermodynamic limit may be taken by counting the number of 
monomino configurations on consecutive lines of length $N$.
In this simple case, the densities $\rho_i$
\footnote{
We suppress the overbar and write $\rho_i$ instead of $\bar{\rho_i}$ from now on.}
can be given as explicit expressions of the activities:
\be
\rho_i = \frac{z_i}{z_1 + \ldots + z_M} \qquad (i=1,\ldots,M) .
\ee
In order to eliminate the normalization constraint, we choose $\rho_1,\ldots,\rho_{M-1}$
as independent densities and normalize the grand-canonical potential via
\be \tilde{\phi}(\mu_1,\ldots,\mu_{M-1}) := \phi(\mu_1,\ldots,\mu_{M-1},0). \ee
Both functions are related via (\ref{form:cons}),
\be \tilde{\phi} \left(\mu_1-\mu_M,\ldots,\mu_{M-1}-\mu_M \right) = 
  \phi(\mu_1,\ldots,\mu_{M}) - \mu_M. \ee
We obtain the entropy as Legendre transform of the normalized grand-canonical potential,
\be 
s(\rho_1,\ldots,\rho_{M-1}) = -\sum_{j=1}^M \rho_j\log \rho_j,
\ee
where $\rho_M=1-\sum^{M-1}_{j=1} \rho_j$.
This coincides with the entropy of a Bernoulli system \cite{Pet83}
with fixed frequencies $p_i=\rho_i$. 
The entropy function is strictly concave and takes its (unique) maximum at the point
$\rho_1 = \cdots = \rho_M = \frac{1}{M}$, with value
\be s_{\rm max}=-\sum_{j=1}^M\frac{1}{M}\log\frac{1}{M}=\log M. \ee
Near its maximum, the entropy curve is a quadratic function of the independent
order parameters.
Its shape near this point is completely determined by the Hessian
\be \left. \frac{\partial^2 s}{\partial \rho_i\partial \rho_j}\right|_{s=s_{\rm max}}=
  -\left. ( \rho_M^{-1}+ \rho_i^{-1}\delta_{ij})\right|_{s=s_{\rm max}}=
  -M(1+\delta_{ij}) \qquad (i,j=1,\ldots,M-1). \ee
The negative Hessian is the elastic tensor. In particular, for $M=2$, we obtain
\be
s(\rho) = \log 2 - 2\left(\rho-\frac{1}{2}\right)^2 +
{\cal{O}}\left(\left(\rho - \frac{1}{2}\right)^4\right)
\ee
which (with $E = 2\rho -1$) conforms to the phasonic expression given in \cite{Hen91}.

Owing to the lack of geometric or other constraints, this class of examples is 
fully controlled by the strong law of large numbers \cite{Bau91}:
With probability one, for a given member of the ensemble, the probability to 
occupy a position with a tile of type $j$ is its frequency, $p_j$.
This even connects the growth with the equilibrium properties, as also observed 
recently for other, more complicated ensembles \cite{JE}.

A first natural step to examine the effect of different tile
geometries as well as of certain packing rules is to allow these in
one coordinate direction. 
Introducing different tile sizes, one of the simplest models is the 1D
monomino-domino model \cite{HL}.

Similar to the previous example, the only constraint of this model is due to
density normalization.
We eliminate this constraint in the same way, i.e. we write
\be {\cal Z}_N(\mu)=\sum_{k=0}^N g^{}_N(k) z^k \ee
where $g_N(k)$ is the number of connected patches with $N$ occupied sites, $k$ of
which being monominoes. 
So, $\mu=\log z$ is the chemical potential of a monomino. 
Obviously, ${\cal Z}_N(\mu)$ fulfils the Fibonacci-type recursion formula
\be 
{\cal Z}_N(\mu)=z {\cal Z}_{N-1}(\mu)+{\cal Z}_{N-2}(\mu)\label{rekursion}
\ee
\bdm    
{\cal Z}_0 := 1 \qquad {\cal Z}_1(\mu)= z. 
\edm
The quotient ${\cal Z}_N(\mu)/{\cal Z}_{N-1}(\mu)$ is a rational function whose limit, 
for $N\to\infty$, exists and can easily be calculated from (\ref{rekursion}) to be 
\be 
\lim_{N\to\infty}\frac{{\cal Z}_N(\mu)}{{\cal Z}_{N-1}(\mu)} = 
\left( \frac{z}{2}+\sqrt{1+\left(\frac{z}{2}\right)^2} \right). \label{grenzwert} 
\ee
The grand-canonical potential per site is defined through
 $\phi(\mu)=\lim_{N\to\infty}\frac{1}{N} 
\log {\cal Z}_N(\mu)$, but can directly be calculated from the quotient (\ref{grenzwert}) as
\be 
\phi(\mu)  = \lim_{N\to\infty}\log\frac{{\cal Z}_N(\mu)}{{\cal Z}_{N-1}(\mu)}
=\arsinh \frac{z}{2}. 
\ee
We denote the monomino density by $\rho_1$, the domino density by $\rho_2$ 
($\rho_1+\rho_2=1$).
Let us now write the entropy curve as a function of the monomino density $\rho_1$. 
The entropy curve can be computed from the grand-canonical potential as
\bea 
s(\rho_1)  &=& \phi(z(\rho_1)) - \rho_1\log z(\rho_1)  \\
&=&(\rho_1+\frac{\rho_2}{2})\log(\rho_1+\frac{\rho_2}{2})-
\rho_1 \log \rho_1 -\frac{\rho_2}{2} \log \frac{\rho_2}{2}, \nonumber 
\eea
where
\bdm 
\rho_1(\mu)  = \frac{\partial \phi(\mu)}{\partial \mu} = \tanh \phi(\mu) = 
\frac{\frac{z}{2}}{\sqrt{1+\left(\frac{z}{2}\right)^2}} \, .
\edm
The entropy function is a strictly concave function with a quadratic maximum 
\be 
s_{\rm max} = \log \tau \approx 0.4812 
\ee
at $\frac{2\rho_1}{\rho_2}=\tau\approx 1.6180$, where $\tau=\frac{1+\sqrt{5}}{2}$
denotes the {\em golden number}
\footnote{
The {\em silver number} $\lambda=1+\sqrt{2}$
appears in the monomino-domino model of monominoes with two different colours.
}.
We obtain an elastic constant $\lambda=\frac{5}{4}\sqrt{5}\approx 2.7951$.

It is possible to derive a closed expression for the grand-canonical partition sum \cite{HL}.
Defining $P_N(x) = i^{-N} {\cal Z}_N( 2 i x)$ transforms the recursion
(\ref{rekursion}) into that of the Chebyshev polynomials of the second kind \cite{AS}
which are
\be P_N(\cos\theta) = \frac{\sin(N+1)\theta}{\sin\theta} \ee
wherefrom one gets
\be {\cal Z}_N(\mu) = \frac{1}{\sqrt{1+\left(\frac{z}{2}\right)^2}} 
\left\{ \ba{c} \cosh \\ \sinh \ea \right\}
\left( (N+1) \; \arsinh \frac{z}{2} \right), \quad \mbox{for } N 
\left\{ \ba{c} \mbox{even} \\ \mbox{odd} \ea \right\}. \ee

\subsubsection*{Packing rules and supertiles}

So far, no further constraint or packing rule has been imposed on the
ensembles. A natural way to do this in one dimension is to exclude
certain strings of consecuting tiles. In many cases, these ensembles
can be directly dealt with by recursive methods as shown above (e.g. the 
$M$-colouring problem in 1D), or the constraints can be eliminated by a 
reformulation of the problem in terms of `supertiles'.
Let us illustrate this for the ensemble of chains in two 
monominoes $a,b$ with the exclusion of $r$ consecuting $a$'s and $s$
consecuting $b$'s.
A moments reflection shows that any allowed chain can (up to boundary tiles) uniquely 
be written in supertiles $A_{i,j}$, $1\le i< r,\, 1\le j< s$, forming
strings of $i$ consecuting $a$'s, followed by $j$ $b$'s. Vice versa,
any sequence in the supertiles translates to an allowed one in $a,b$.

Similarly, the ensemble of monomino chains with tiles $a,b$ and exclusion of $bb$ 
is equivalent to the monomino-domino model mentioned above. 
We give the entropy for this ensemble via its `supertile' formulation
with polyominoes $A=a$ and $B=ab$, starting from a Bernoulli ensemble rather than 
taking different tile geometries of the supertiles into account.
The frequencies of the two models are related according to
\be p_a = \frac{p_A+p_B}{p_A+2 p_B}, \qquad  p_b = \frac{p_B}{p_A+2 p_B}, \ee
\bdm p_A = 1 - \frac{p_b}{p_a}, \qquad p_B = \frac{p_b}{p_a}. \label{umrech}\edm
It is obvious from here that the possible values of $p_a, p_b$ are 
restricted to
\be \frac{1}{2} \le p_a,  \qquad 0 \le p_b \le p_a \le 1.\ee
As the entropy per supertile is given by
\be \tilde{s}(p_A,p_B) = -p_A\log p_A -p_B\log p_B, \ee
the entropy per (small) monomino follows as
\bea s(p_a,p_b) &=& \left(1+\frac{p_b}{p_a}\right)^{-1} 
     \tilde{s}(p_A(p_a,p_b),p_B(p_a,p_b)) \nonumber\\
  & = & p_a \log p_a - p_b \log p_b -(p_a-p_b)\log(p_a-p_b). \eea
Taking $x=2 p_a -1$ as parameter, $0 \le x \le 1$, the entropy curve is 
exactly the curve of the monomino-domino model.
The maximum occurs at $p_a=\frac{2+\tau}{5} \approx 0.7326$.
We obtain an elastic constant $\lambda=5 \sqrt{5} \approx 11.1803$.

However, already in 1D, more complex examples may be considered. An interesting
family is provided by the ensemble of square-free words in $n$ letters, 
$n \ge 3$, which is known to display positive entropy. In fact, the value
\be \tilde{s}(n)=\arcosh\left(\frac{n-1}{2}\right) \ee
gives (for $n \ge 4$) a reasonable lower bound of it, which is asymptotically 
exact, but the determination of the exact value is still an open problem \cite{BEG}.

\subsection{Entropy and symmetry}

In this section we consider effects of the symmetries of a tiling 
ensemble on its entropy function.
It should be stressed again 
that the results below are derived using the grand-canonical formalism.
All physical quantities are averages over the whole tiling ensemble.
In most situations of physical relevance, however, the ensemble averages
coincide with the corresponding quantities of a typical single tiling, and
all results translate to the canonical ensemble as well.
This will be the case for all examples discussed in this article.

As a symmetry of a random tiling ensemble, we define every linear bijection
$S_{\rho}^{}$ on the density parameters that leaves the entropy function invariant.
We write
\begin{equation}
S_{\rho}^{}: \rho_i^{} \mapsto \tilde{\rho_i^{}} =\sum_{j=1}^M S_{ij}^{}\rho_j^{}
\qquad (i=1,\ldots,M).
\label{form:sro}
\end{equation}
Owing to the normalization constraint, the restriction of $S_{\rho}^{}$ to the subspace
of {\em independent} density parameters where the entropy function is defined leads 
in general to an {\em affine} bijection.

Examples are obtained from bijections on the tiling ensemble which transform
densities like (\ref{form:sro}).
They are symmetries of the random tiling ensemble if they respect the thermodynamic
limit, i.e. if they induce a bijection on the set of $\Lambda$-patches, for each
$\Lambda$ of a suitably chosen limit sequence $\{\Lambda\}$.
The simplest examples are colour symmetries, which permute equal prototiles
of different colour.
Moreover, rotations and reflections on the tiling ensemble, together with a 
limit sequence of circular sets, lead to {\em geometric} symmetries.
Let us denote every symmetry which is not of the former type as {\em hidden} symmetry.
We will discuss a random tiling with a hidden symmetry below.

In general, every  linear bijection on the densities can as well be represented in 
the vector space of all chemical potentials;
as can be verified from (\ref{form:Dichte}), a transformation $S_{\rho}^{}$ induces
the adjoint mapping
\be S_{\mu}: \mu_i^{} \mapsto \tilde{\mu_i^{}} =\sum_{j=1}^M \left(S^*\right)_{ij} \mu_j
\qquad (i=1,\ldots,M),
\label{form:musymm} \ee
where $S^* = (S^{-1})^t$ means the matrix inverse and transpose of $S$.
In particular, the invariance of the grand-canonical potential under symmetry operations 
(\ref{form:musymm}) imposes symmetries of the (mean) density vector field in the space
of chemical potentials,
\be \tilde{\rho}(\mu) = \rho(\tilde{\mu}). \label{form:dtraf}\ee

We conclude immediately that symmetry related tiles having equal chemical 
potential occur with the {\em same} (mean) density.
Note that the point of maximum entropy $(\mu = 0)$ is a 
{\em fixed point} under any given symmetry which means that symmetries constrain
possible density parameters at the entropy maximum.
This is precisely the statement of the so-called first random tiling hypothesis:
\begin{itemize}
\item The point of maximum entropy is a point of maximum symmetry.
\end{itemize}
For random tilings with a height representation, maximum symmetry implies zero
phason strain, $E \equiv 0$ \cite{Hen91}, so this is contained.
In our setting, the conlusion above implies:
\begin{itemize}
\item At the point of maximum entropy, symmetry related tiles occur with 
equal (mean) density.
\end{itemize}
We can further conclude,
as can be read off from (\ref{form:dtraf}), the density vector $\rho$, at the point
of maximum entropy, has only components in direction of the trivial one-dimensional
representations of the symmetry group.
Thus, the symmetry determines the entropy maximum completely, if the representation 
of the symmetry group on the vector space of {\em independent}
density parameters is irreducible (and nontrivial).
We will give examples of that kind later on.

Let us now focus on the transformation behaviour of the entropy function 
(\ref{form:entropie}).
This is done most easily within a symmetry-adapted parametrization.
By an affine coordinate transformation, it is always possible to introduce new 
independent density parameters such that the new origin is the point of maximum entropy, 
and the symmetries are represented by linear, orthogonal transformations.
In many cases, the new density parameters have a geometric interpretation in form of 
densities of supertiles.
Let us call these new parameters $r$ and assume that we are in such a frame.

The entropy is invariant under symmetry transformations,
\be s(r) = s(\tilde{r}) .\ee
At its maximum, the entropy expansion can be written in terms of invariants of the 
given symmetry group
\footnote{
For the {\em quasicrystal} entropy as a function of perp 
strain, this kind of arguing was used in \cite{Lu}.
}.
The form of the second order terms is determined by the Hessian $H$ of the entropy
(the matrix of its second derivatives).
The symmetries induce a transformation behaviour of the form
\be \left.H\right|_{r} = (S_{r}^{})^t \left.H\right|_{\tilde{r}} S_{r}^{}.\ee
Since $S_{r}$ is an orthogonal transformation, it {\em commutes} with $H$ at the point of 
maximum entropy,
\be \left. H \right|_o  S_{r} = S_{r} \left. H \right|_o. \ee
According to Schur's lemma \cite{Ham}, $\left. H \right|_o$ acts trivially on the irreducible 
subspaces of the symmetry group and can thus be written as a linear combination of
projectors $P^{(i)}$ onto the irreducible components,
\be \left. H \right|_o = -\sum \lambda_i P^{(i)}. \ee
We refer to the $\lambda_i$ as {\em elastic constants} of the given random tiling.
This is more appropriate than the use of the eigenvalues since they 
are no longer independent in the presence of symmetries.
The quadratic term of the entropy expansion is then of the form
\be s_2(r,r)=-\frac{1}{2}\sum \lambda_iI^{(i)}(r,r), \ee
where the $I^{(i)}(r,r)$ are the quadratic invariants defined by
\be I^{(i)}(r,r)=\quad<r,P^{(i)}r>, \ee
and $<\cdot,\cdot>$ denotes the scalar product.
We will give examples later on.

Now, the discussion on the relation of entropy and symmetry can be extended
even further.
Above, we have always taken a set of tiles as basic
elements to construct the tiling.
This is, however, not the only possibility. 
Instead, tilings can also be seen as built from bonds, vertex configurations, 
patches of a given volume or number of tiles, or the like. 
Adopting this point of view, we can assign chemical potentials and densities 
to patches or bonds rather than to the prototiles.
While the number of basic elements and packing rules may increase quite rapidly
this way, the above arguments can nevertheless be repeated without
change, and we conclude that ensemble densities of symmetry related patches
{\em of any size} are equal at the point of maximum entropy. 
If the density parameters of a typical tiling of the ensemble coincide
with their ensemble averages, then:
\begin{itemize}
\item Entropically stabilized random tilings are (with probability
one) {\em locally indistinguishable}\footnote{An introduction to equivalence 
concepts of tilings is given in \cite{BS95}.}
from their images under symmetry transformations.
\end{itemize}
This is the main result of this section.
Note that, also for tilings with a height representation, this is a much
stronger statement than the demand $E \equiv 0$ on the phason strain,
since different densities of symmetry related patches may be
compatible with the same value for the phason strain.  

\subsection{Examples in two dimensions}

\subsubsection{The rhombus tiling --
a domino tiling of Kasteleyn type with height representation}

The rhombus random tiling \cite{E1,BH} is the ensemble of all coverings of the plane 
with $60^{\circ}$-rhombi.
It is also called a lozenge tiling.
It is a domino tiling on the triangular lattice, equivalent to the fully packed 
dimer model on the hexagonal lattice \cite{K2,W}.
We discuss it here since it is the prototype of a random tiling possessing a 
height representation, which allows us to demonstrate the equivalence of our
approach to the conventional one, for this model.
In the following, we focus on the symmetries of the tiling and give a simple derivation 
of the entropy.
In the appendix, we relate the elastic constant given by Henley \cite{Hen88,Hen91} to our 
result.

The symmetry of the random tiling ensemble which transforms rhombus tilings into
other allowed ones is $D_6 = D_3 \times C_2$. 
For undecorated tiles, however, the inversion symmetry of the prototiles
already enforces $C_2$ as (minimal) symmetry for any element of the
tiling ensemble, thus $S_3 \simeq D_3$ remains as relevant geometric symmetry in the
space of density parameters. 
There are no hidden symmetries in this case.
Since all prototiles are related by rotation, for the point of maximal entropy we 
immediately conclude $\rho_1 = \rho_2 = \rho_3 = \frac{1}{3}$. 

The filling constraint $\rho_1+\rho_2+\rho_3=1$ reduces the number of independent 
variables to two. 
As the symmetry acts irreducibly on the remaining two-dimensional
subspace, the symmetry of the entropy function must belong
to a two-dimensional irreducible representation of $D_3$. 
It is most convenient to represent the phase space as the set of points inside an 
equilateral triangle, see figure \ref{fig:dreieck}, this way obtaining a
handy parametrization which we call its phase diagram.
\Bild{4}{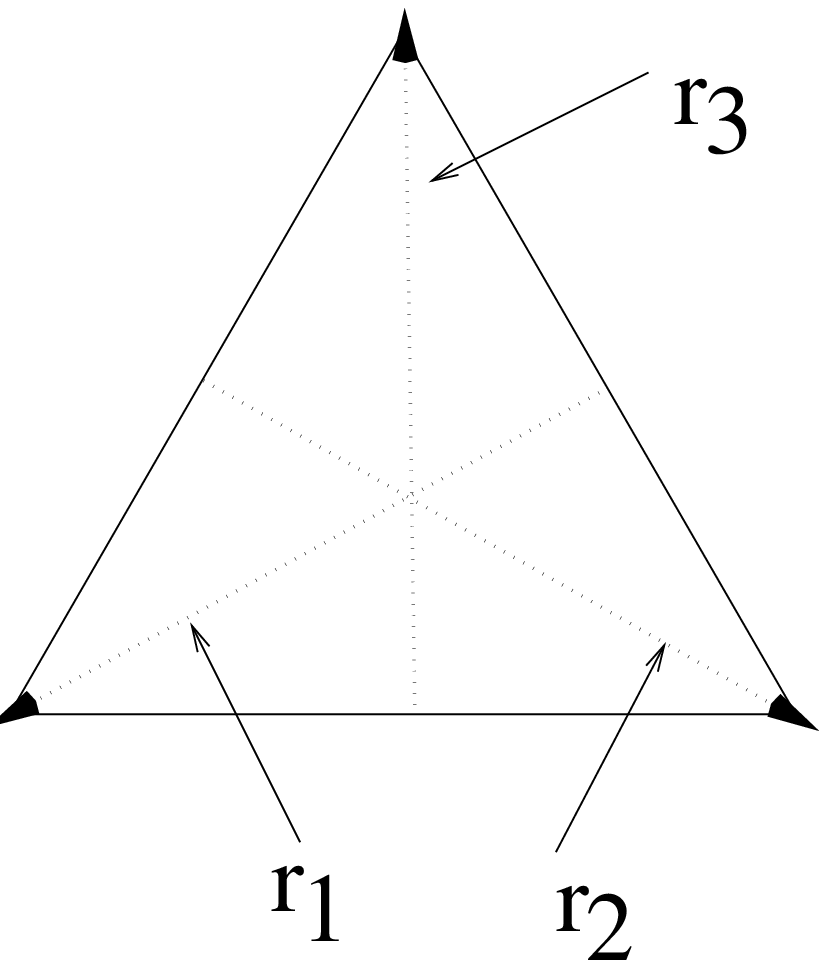}{The phase space of the rhombus tiling is an equilateral triangle.}
{fig:dreieck}
We choose its center as the origin and the vectors pointing to the corners of the triangle
as unit vectors of the different tile densities.
The center corresponds to the point of maximal entropy, and each point 
$r=(r_1,r_2,r_3)$ corresponds to an ensemble with densities $\rho_i=r_i+\frac{1}{3}$.
The symmetries of the entropy function are then reflected in the symmetries of 
the triangle.
The point of maximum entropy is the only one with $D_3$ symmetry. 
Along lines with equal density of two different tiles there is reflection symmetry.

The grand-canonical potential per rhombus as a function of the different tile potentials 
$\mu_i=\log z_i$ (with $i=1,2,3$) reads \cite{W}
\bea \phi(\mu_1, \mu_2, \mu_3) &=&\frac{1}{8\pi^2}\int\limits_0^{2 \pi}\int\limits_0^{2 \pi} 
\log \det \lambda (\varphi_1,\varphi_2) d\varphi_1 d\varphi_2, \label{free}\\
  &=& \frac{1}{4\pi}\int\limits_0^{2 \pi} \log \left( \max \left\{z_1^2,
z_2^2+z_3^2-2z_2z_3\cos\varphi\right\}\right) d\varphi, \nonumber
\eea
where the determinant is given by
\bdm
\det \lambda (\varphi_1,\varphi_2) = z_1^2 + z_2^2 + z_3^2 + 2 z_1 z_2 \cos\varphi_1
+ 2 z_2 z_3 \cos\varphi_2+ 2 z_3 z_1 \cos(\varphi_1-\varphi_2).
\edm
This result was obtained by taking rhombohedral patches of increasing size, together with
{\em periodic} boundary conditions.
The result applies nevertheless to the corresponding limit with free boundary conditions,
as can be seen as follows.
The rhombus tiling can be regarded as a limiting case of the dart-rhombus tiling described below,
where some chemical potentials are set to $-\infty$.
This operation commutes with the {\em free} thermodynamic limit \cite{LW72}, yielding the same 
result as in (\ref{free}).

We introduce the normalization constraint by setting
\be \tilde{\phi}(\mu_1, \mu_2) := \phi(\mu_1, \mu_2, 0) . \ee
The relation between $\tilde{\phi}$ and $\phi$ is given by
\be \tilde{\phi}\left(\mu_1-\mu_3, \mu_2-\mu_3 \right) = 
    \phi(\mu_1, \mu_2, \mu_3) - \mu_3, 
\ee
\bdm 
  \tilde{\rho}_1 \left(\mu_1-\mu_3, \mu_2-\mu_3 \right) = 
\rho_1(\mu_1, \mu_2, \mu_3), \qquad
  \tilde{\rho}_2 \left(\mu_1-\mu_3, \mu_2-\mu_3 \right) = 
\rho_2(\mu_1, \mu_2, \mu_3).  \edm
Let us determine the entropy curve as a function of the independent densities
$\tilde{\rho}_1=\rho_1$ and $\tilde{\rho}_2=\rho_2$.
From the normalized grand-canonical potential $\tilde{\phi}$ we obtain
\be
z_1(\tilde{\rho_1},\tilde{\rho_2})=\frac{\sin\pi\tilde{\rho_1}}
  {\sin\pi(\tilde{\rho_1}+\tilde{\rho_2})}, \qquad 
z_2(\tilde{\rho_1},\tilde{\rho_2})=\frac{\sin\pi\tilde{\rho_2}}
 {\sin\pi(\tilde{\rho_1}+\tilde{\rho_2})}. \ee
The expression for the entropy follows as
\be
s(\tilde{\rho_1},\tilde{\rho_2}) = -\int_0^{\tilde{\rho_1}} 
  \log z_1(\tilde{\rho},\tilde{\rho_2}) d\tilde{\rho} =
  -\int_0^{\tilde{\rho_1}} \log \frac{\sin\pi\tilde{\rho}}
 {\sin\pi(\tilde{\rho}+\tilde{\rho_2})} d\tilde{\rho}.
\ee
At the point of maximum entropy we find
\be s_{\rm max} = \frac{1}{\pi}\sum_{k=1}^{\infty} \frac{\sin
\frac{\pi}{3} k}{k^2}\approx 0.3230, \ee
which is considerably smaller than the value 
$\frac{1}{\pi} \sum_{k=0}^\infty \frac{(-1)^k}{(2k+1)^2} \approx 0.583$ 
of the square lattice domino problem \cite{K}.

As the irreducible representation of the symmetry group is two-dimensional, we expect
to have a single elastic constant.
In fact, for the quadratic term in the entropy expansion we find
\be s_2(r,r)=
 - \frac{1}{2}\cdot\frac{\pi}{\sqrt{3}}\cdot 2(r_1^2+r_1r_2+r_2^2). \label{form:rhombquad}\ee
On the other hand, the entropy function is defined on the subspace complementary 
to the vector $(1,1,1)^t$.
This yields
\be I(r,r)=2(r_1^2+r_1r_2+r_2^2). \ee
Thus we find an elastic constant of $\lambda=\frac{\pi}{\sqrt{3}}$.
In the appendix we relate this constant to the constant given by Henley \cite{Hen88,Hen91} 
which was defined from the height representation of this model.

Note that there is a phase transition at points where two densities approach zero.
In the above parametrization, this corresponds to the corners of the triangle.
At these points, the covariance matrix shows a square-root singularity \cite{BH},
indicating a phase transition of Kasteleyn type \cite{N}.

We give an expression of the entropy along a line of reflection symmetry:
in the regime $\rho_1=\rho_2$, we have
\be s(\rho_1)= \frac{2}{\pi} \int_0^{\pi\rho_1}\log 2\cos x dx. \ee
This function is plotted in figure \ref{fig:rhombent}.
The elastic constant is $\lambda=\frac{\pi}{\sqrt3}$.
\Bild{6}{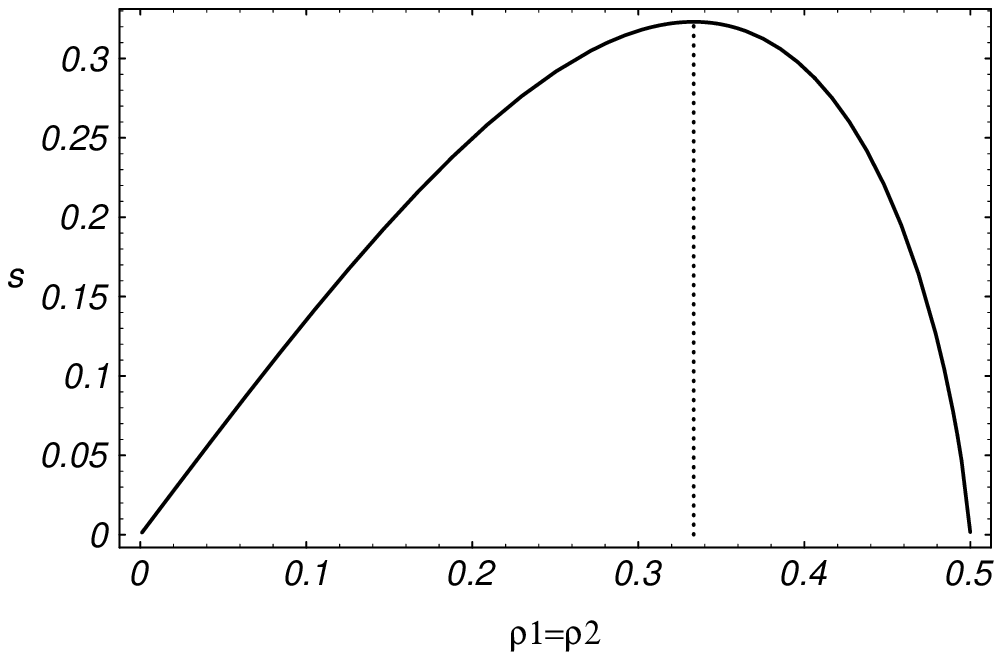}{Entropy of the rhombus random tiling in the
regime $\rho_1=\rho_2$.}{fig:rhombent}

\subsubsection{The dart-rhombus tiling -- 
a domino tiling of Onsager type with hidden symmetry}

\begin{figure}[htb]
  \begin{center}
  \begin{minipage}{13cm}
    \leavevmode
    \epsfig{file=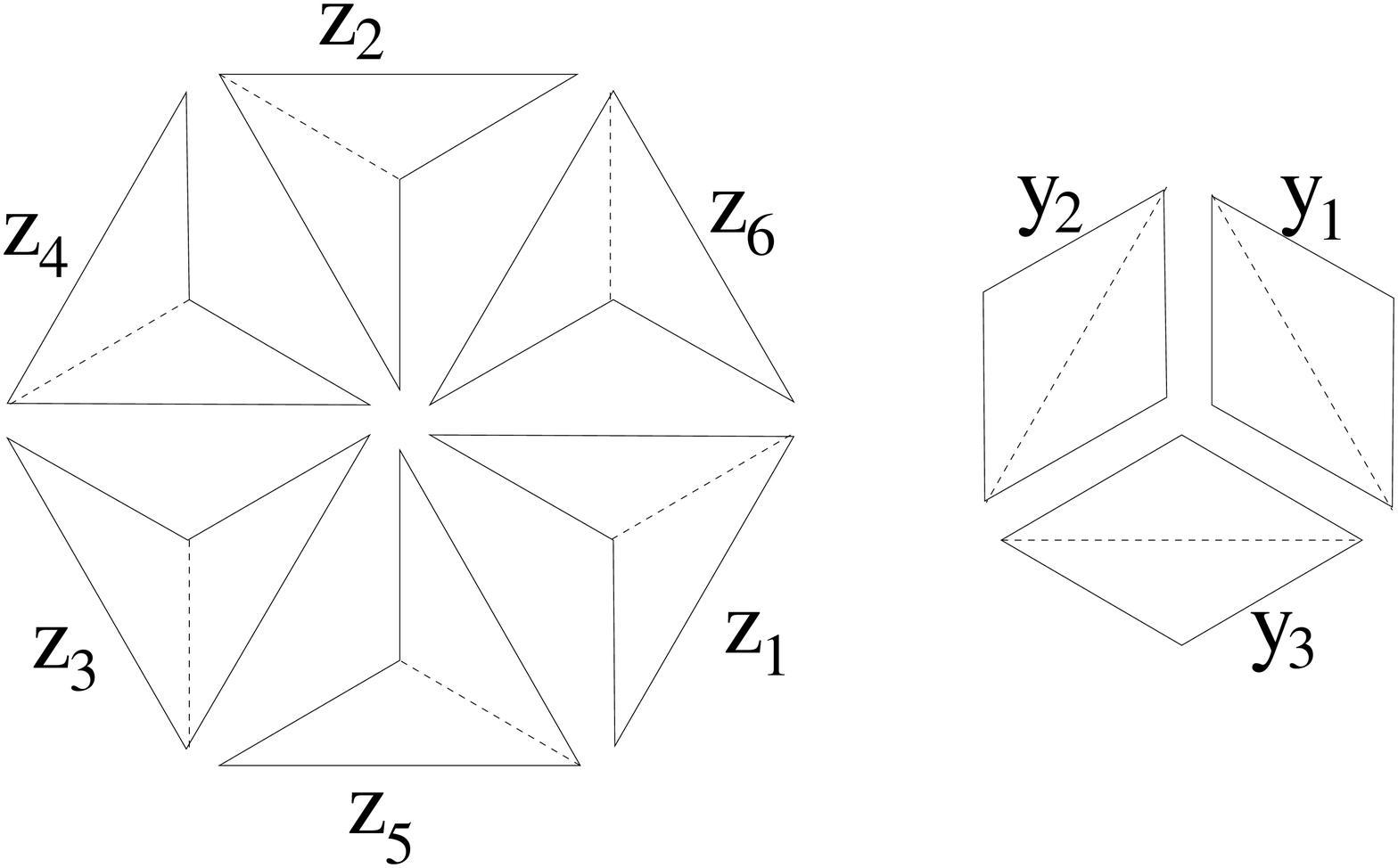,height=4cm}
    \hfill
    \epsfig{file=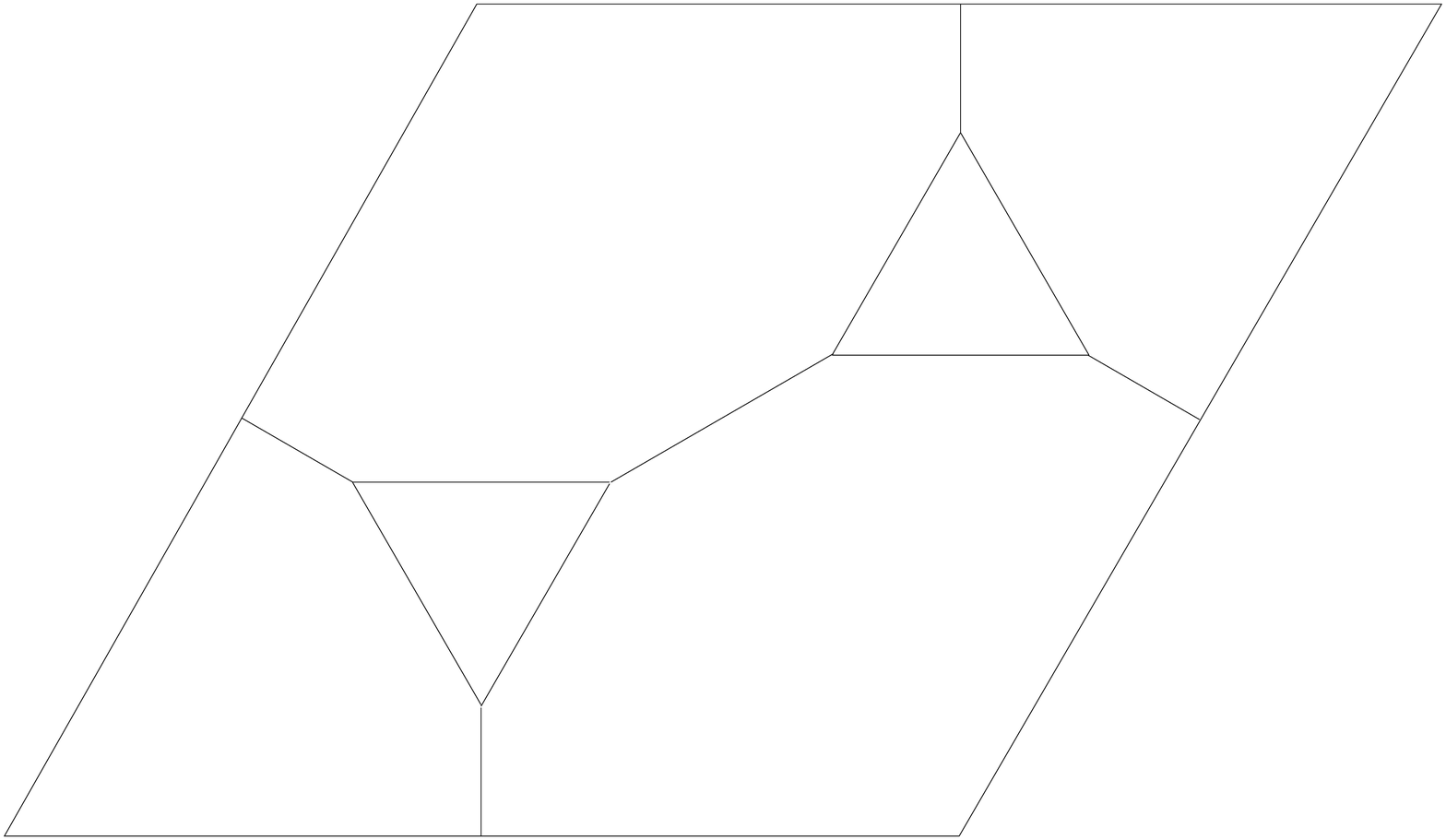,height=3.6cm}  
    \caption{{\small Prototiles of the dart-rhombus tiling and fundamental cell of the 
Fisher lattice.}}
    \label{fig:darttile}
  \end{minipage}
   \end{center}
\end{figure}

\begin{figure}[htb]
  \begin{center}
   \begin{minipage}{6cm}
     \epsfig{file=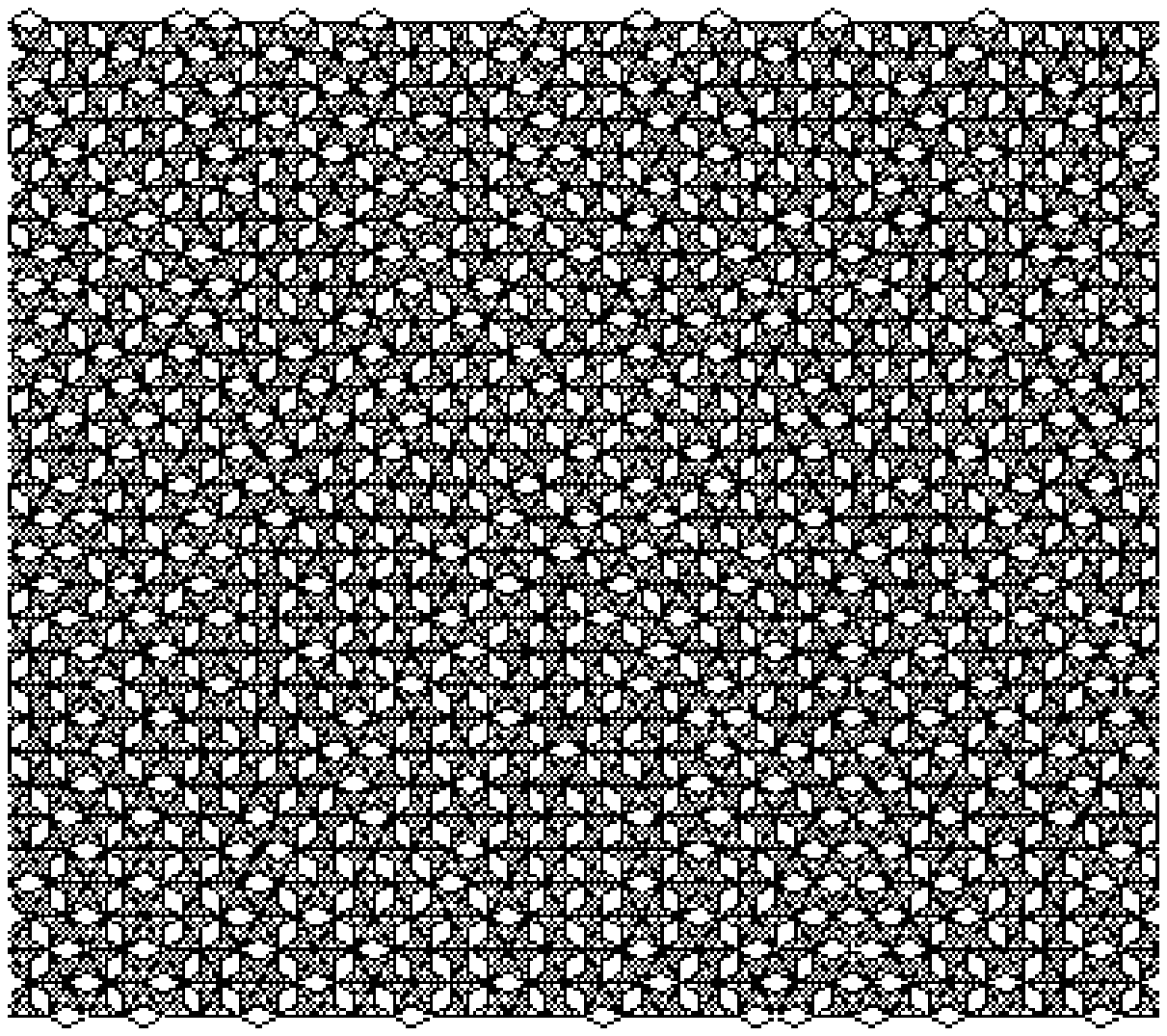,height=6cm}
   \end{minipage}
   \hfill
   \begin{minipage}{9cm}
    \centerline{
      \begin{minipage}{2.9cm}
        \centerline{\epsfysize=2.9cm \epsfbox{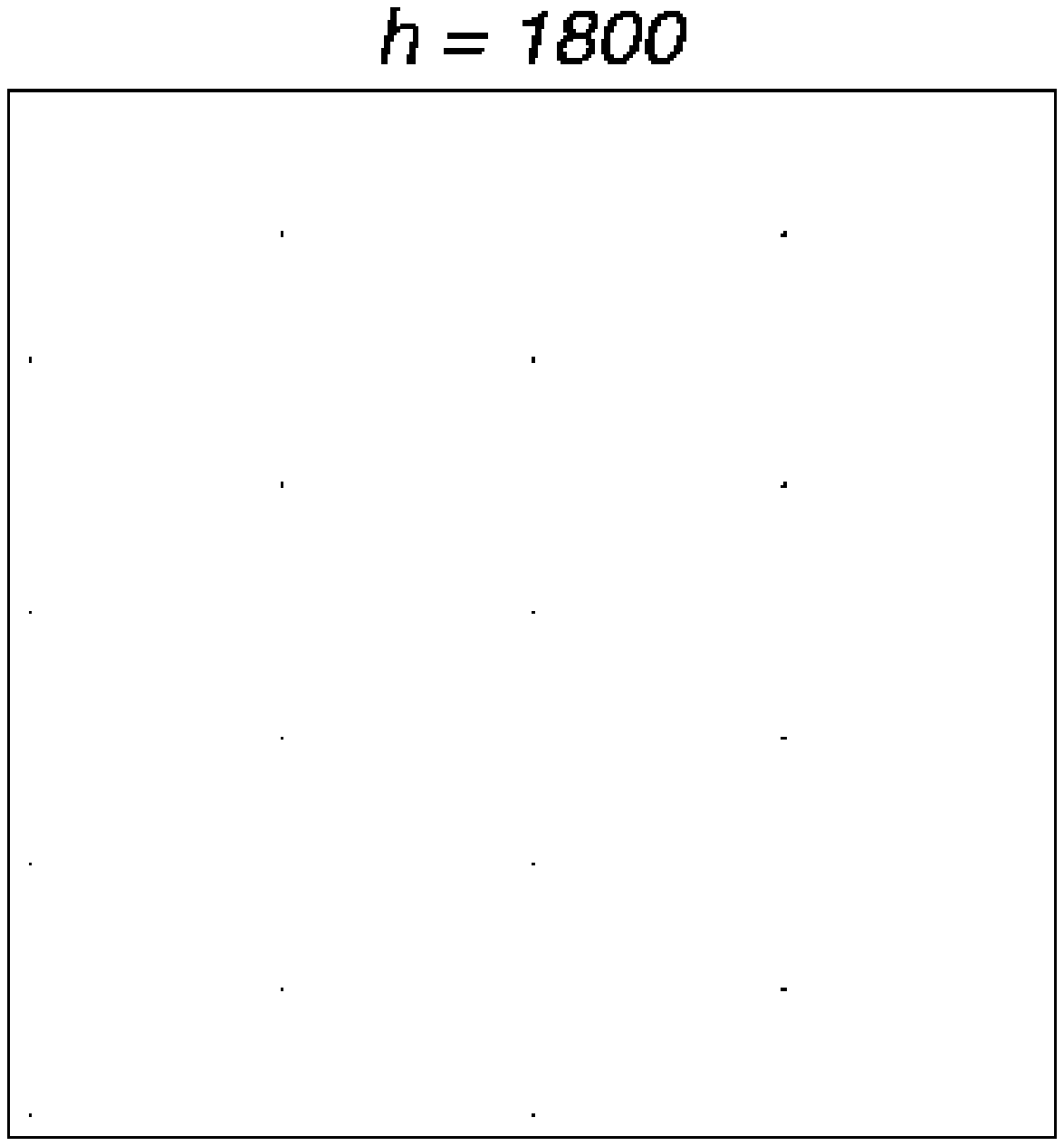}}
      \end{minipage}
      \hspace{1mm}
      \begin{minipage}{2.9cm}
        \centerline{\epsfysize=2.9cm \epsfbox{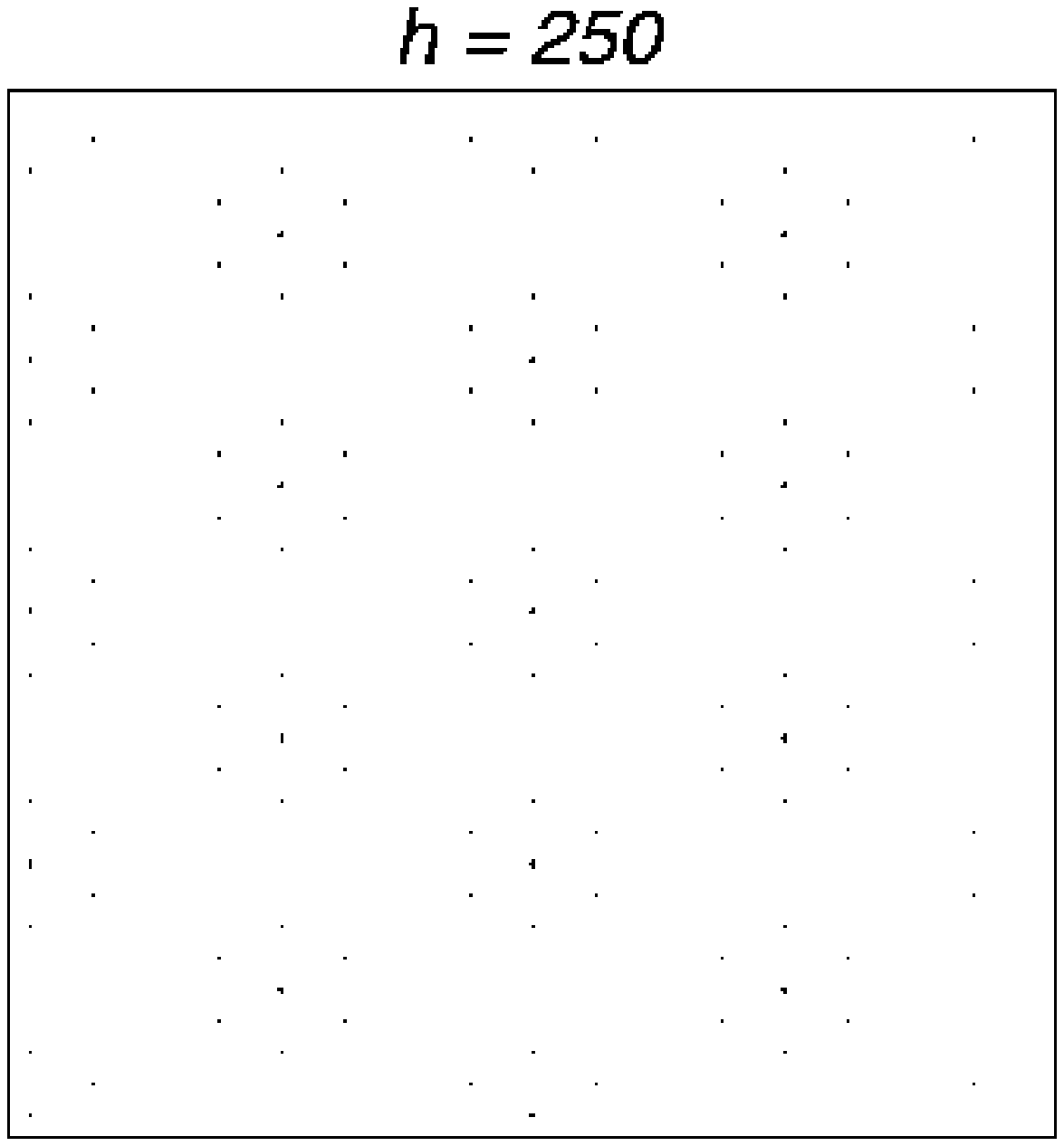}}
      \end{minipage}
      \hspace{1mm}
      \begin{minipage}{2.9cm}
        \centerline{\epsfysize=2.9cm \epsfbox{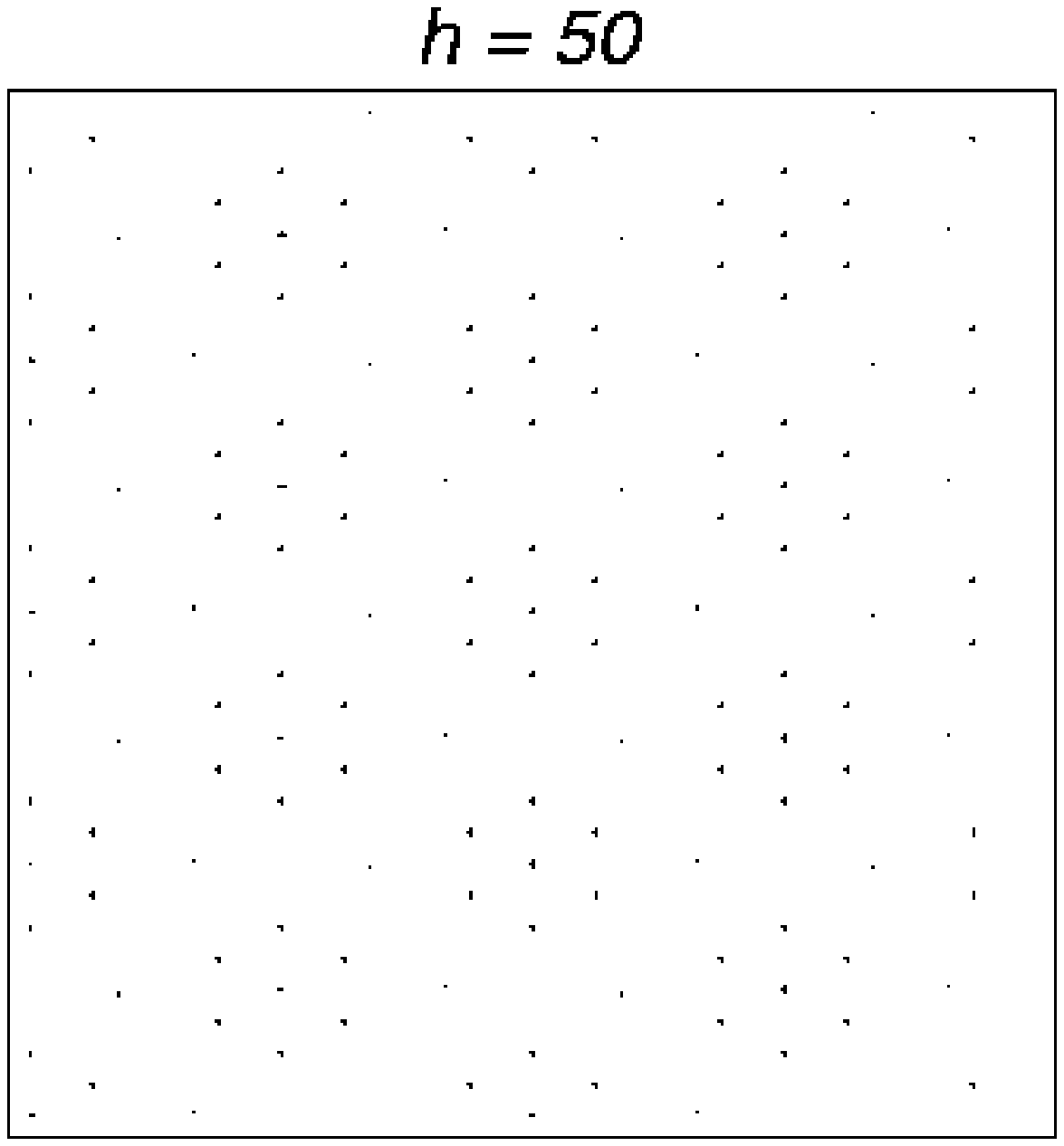}}
      \end{minipage}}
     \centerline{
      \begin{minipage}{2.9cm}
        \centerline{\epsfysize=2.9cm \epsfbox{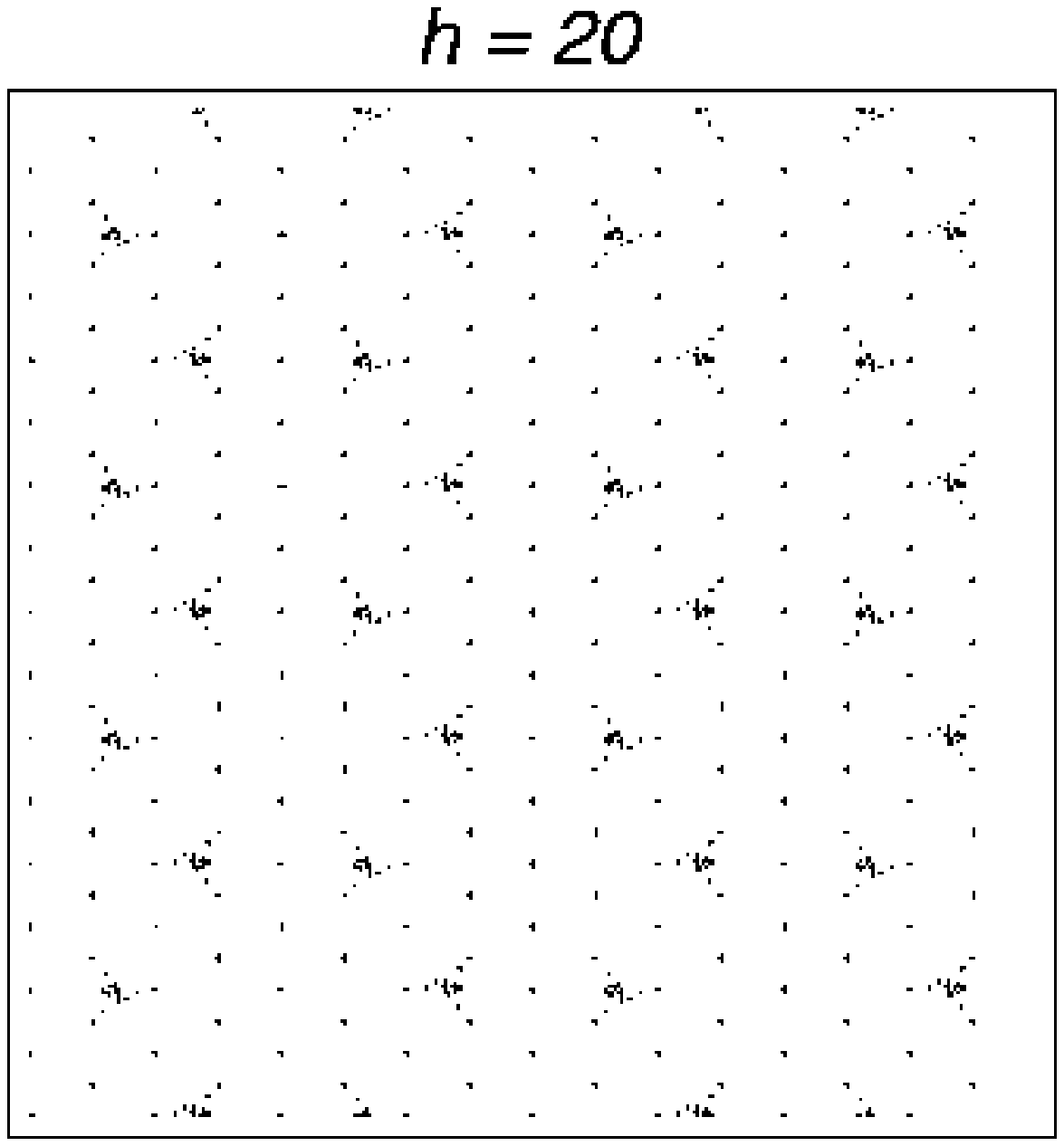}}
      \end{minipage}
      \hspace{1mm}
      \begin{minipage}{2.9cm}
%        \centerline{\epsfysize=2.9cm \epsfbox{pic1/Fourier14.eps}}
        \centerline{\epsfysize=2.9cm \epsfbox{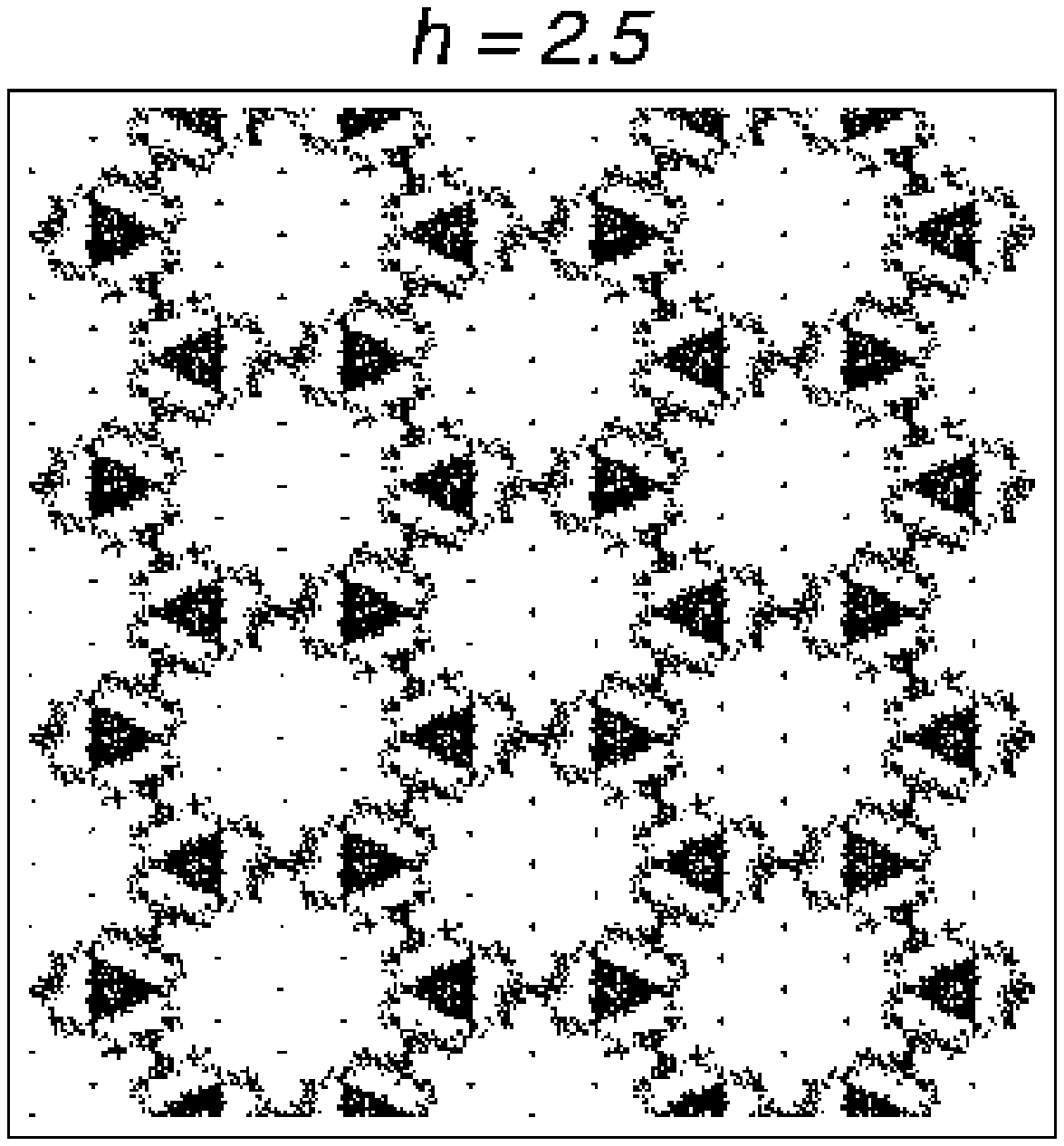}}
      \end{minipage}
      \hspace{1mm}
      \begin{minipage}{2.9cm}
%       \centerline{\epsfysize=2.9cm \epsfbox{pic1/Fourier14.eps}}
        \centerline{\epsfysize=2.9cm \epsfbox{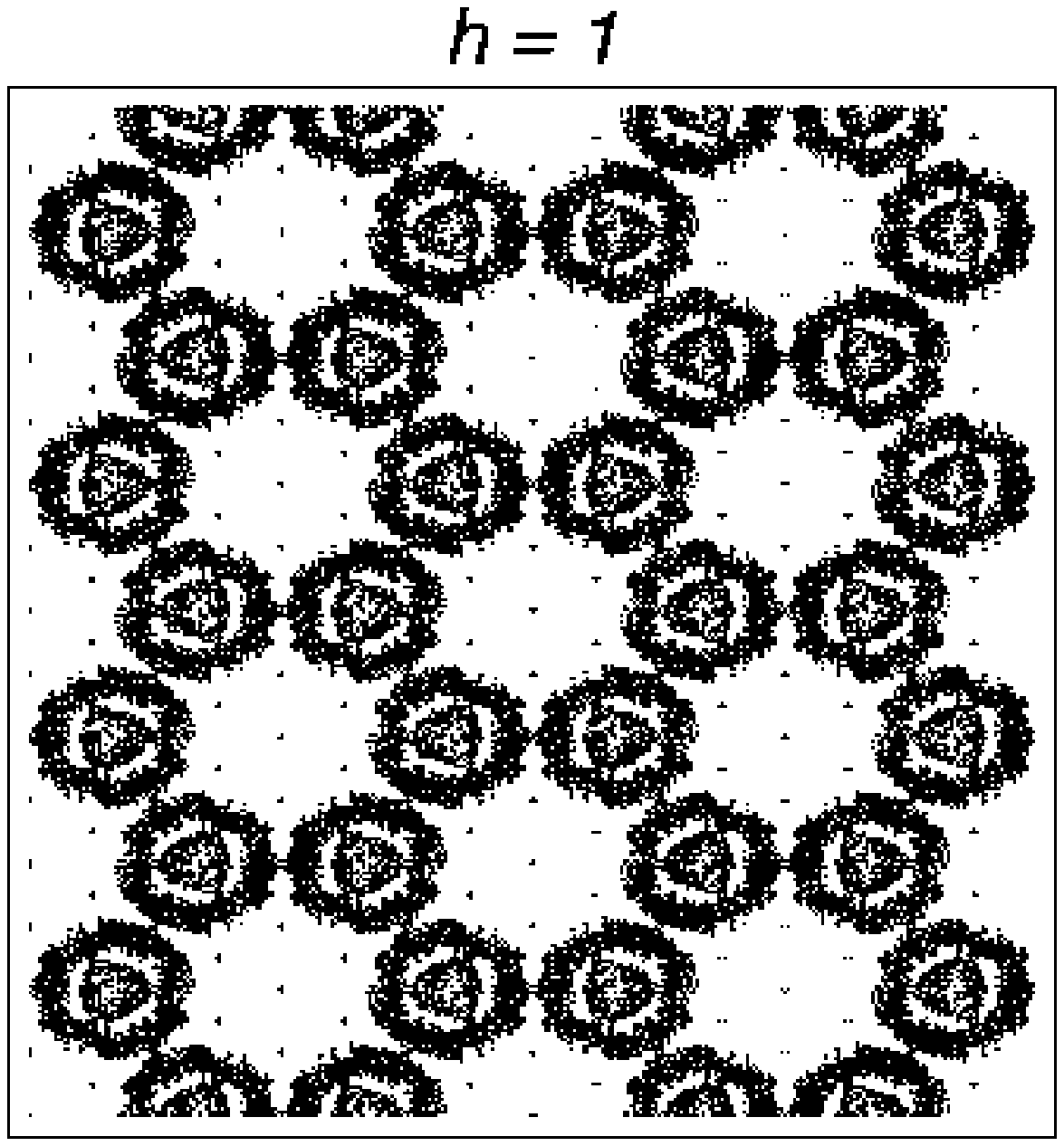}}
      \end{minipage}}
     \end{minipage}
  \end{center}
\caption{{\small Minimal rhombi density ($\rho=0.33$)}}
\label{fig:minimum}
\end{figure}

\begin{figure}[htb]
  \begin{center}
   \begin{minipage}{6cm}
     \epsfig{file=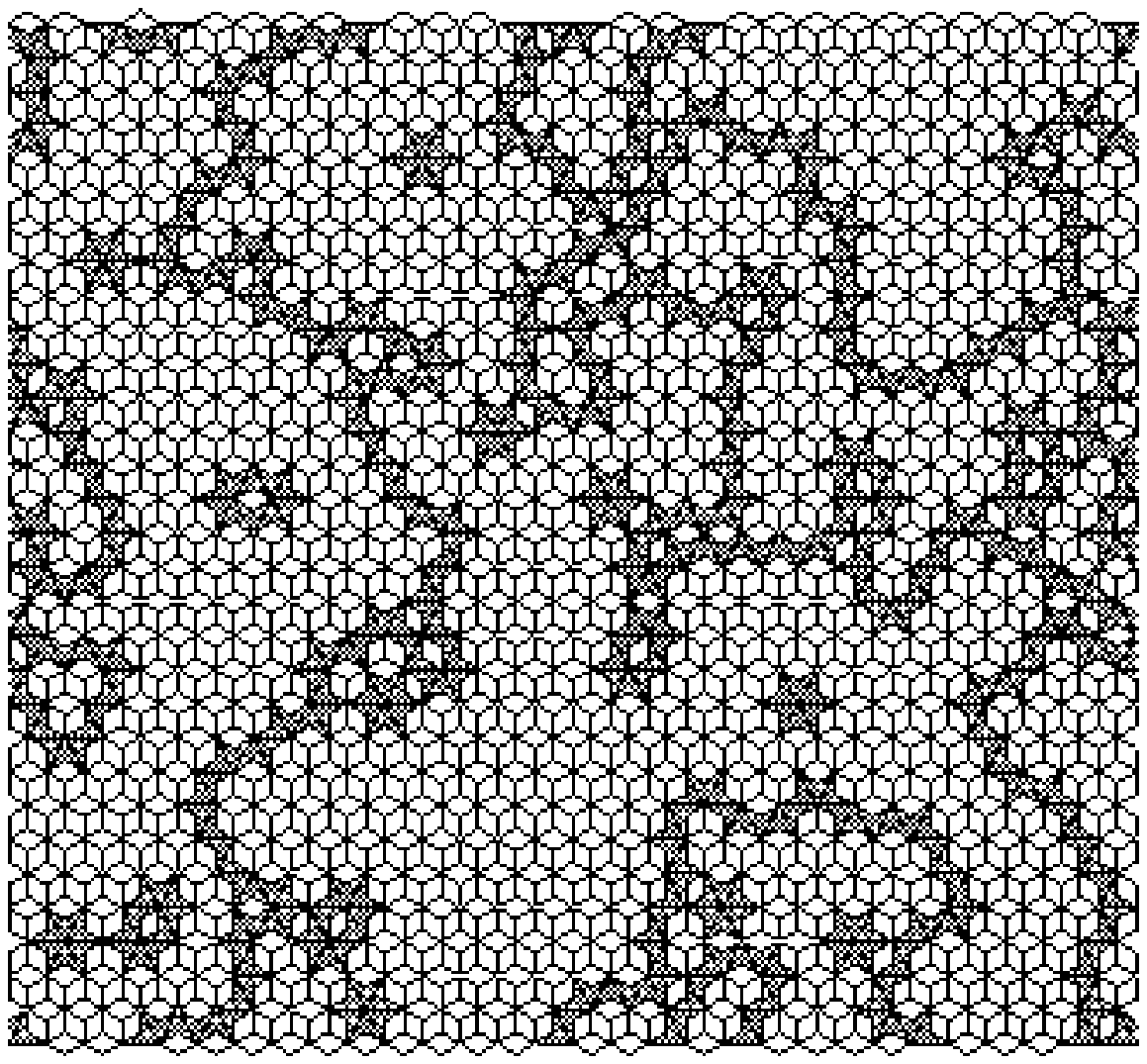,height=6cm}
   \end{minipage}
   \hfill
   \begin{minipage}{9cm}
    \centerline{
      \begin{minipage}{2.9cm}
        \centerline{\epsfysize=2.9cm \epsfbox{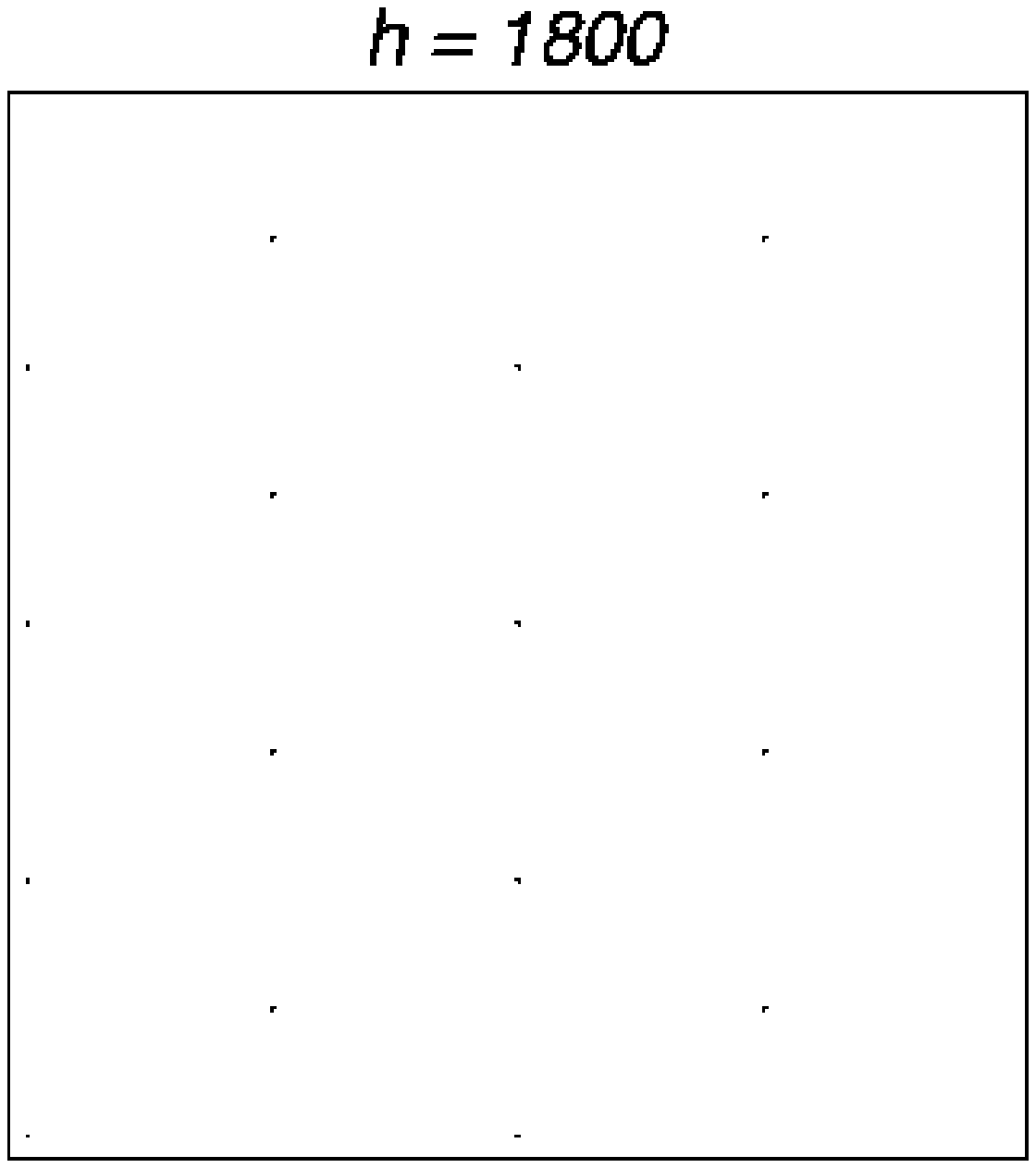}}
      \end{minipage}
      \hspace{1mm}
      \begin{minipage}{2.9cm}
        \centerline{\epsfysize=2.9cm \epsfbox{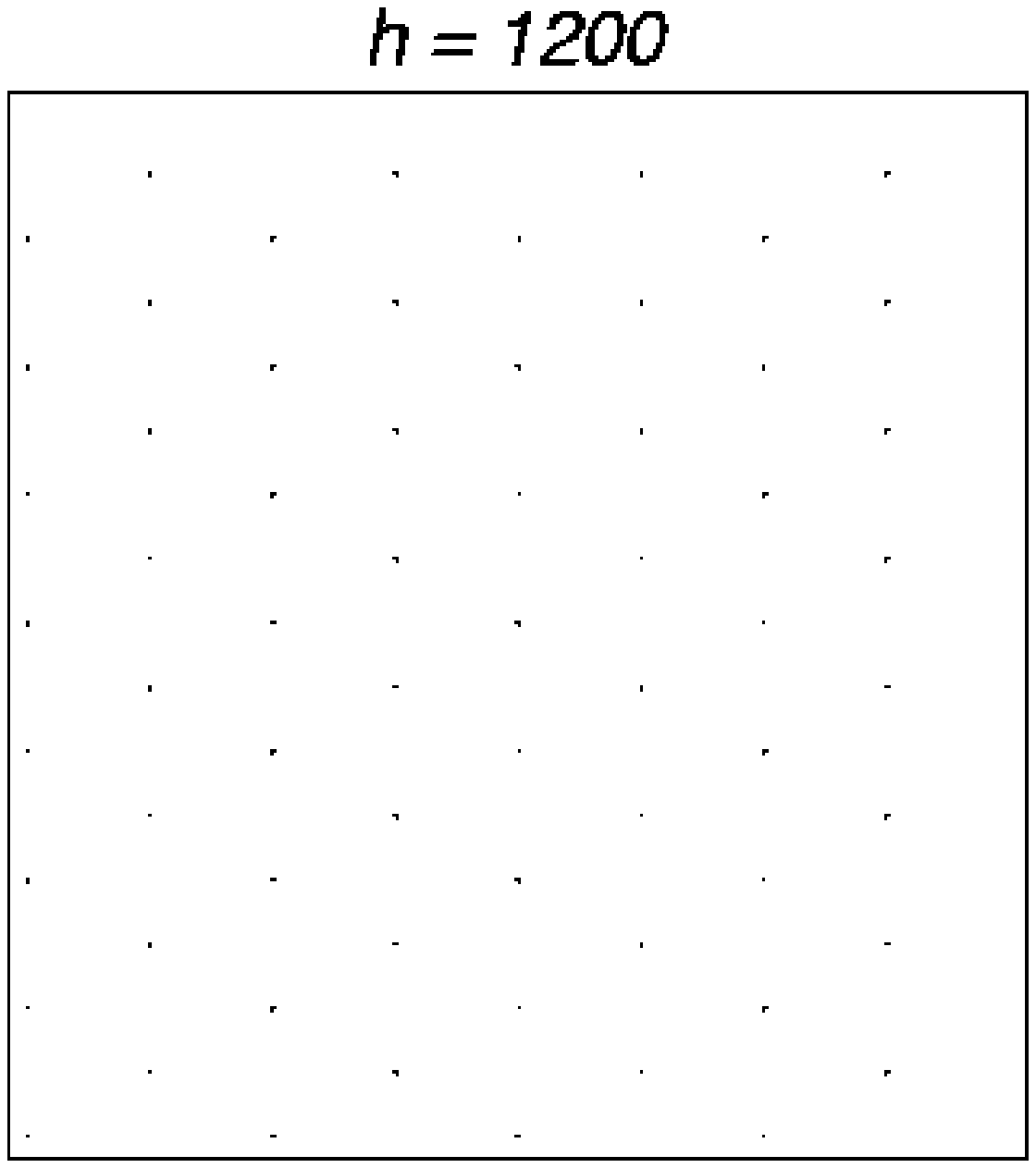}}
      \end{minipage}
      \hspace{1mm}
      \begin{minipage}{2.9cm}
        \centerline{\epsfysize=2.9cm \epsfbox{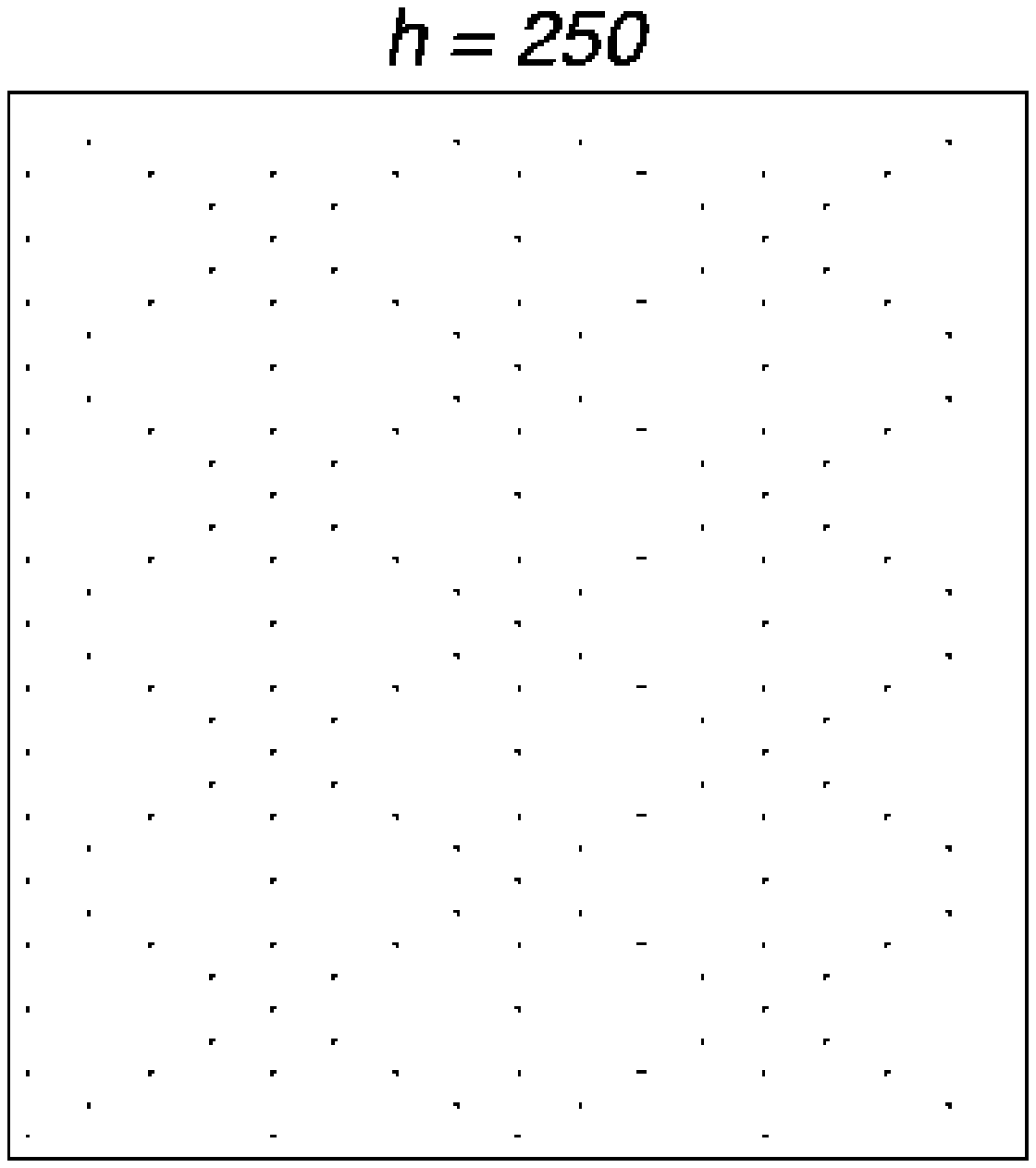}}
      \end{minipage}}
     \centerline{
      \begin{minipage}{2.9cm}
        \centerline{\epsfysize=2.9cm \epsfbox{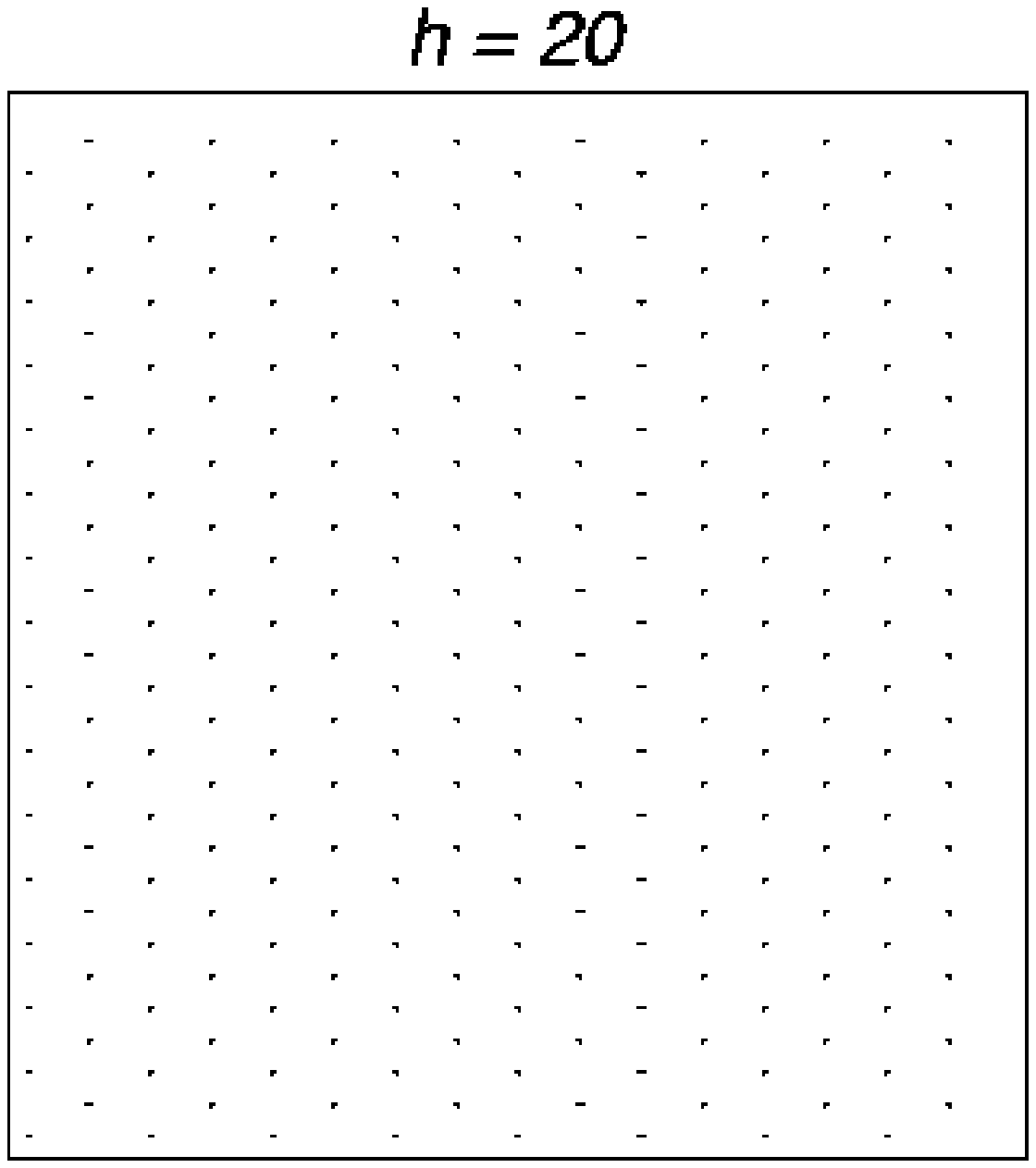}}
      \end{minipage}
      \hspace{1mm}
      \begin{minipage}{2.9cm}
%        \centerline{\epsfysize=2.9cm \epsfbox{pic2/Fourier24.eps}}
        \centerline{\epsfysize=2.9cm \epsfbox{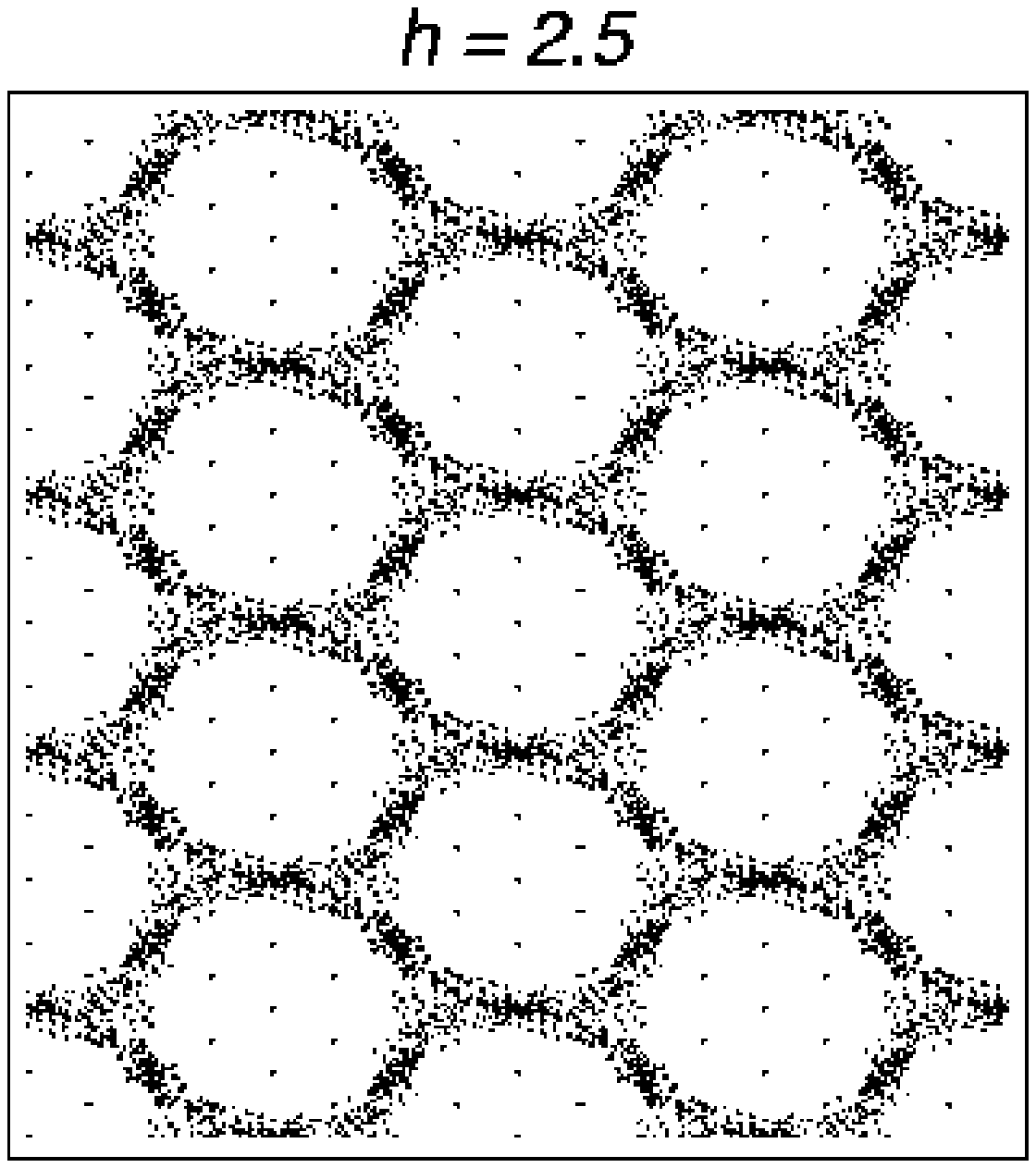}}
      \end{minipage}
      \hspace{1mm}
      \begin{minipage}{2.9cm}
%         \centerline{\epsfysize=2.9cm \epsfbox{pic2/Fourier24.eps}}
        \centerline{\epsfysize=2.9cm \epsfbox{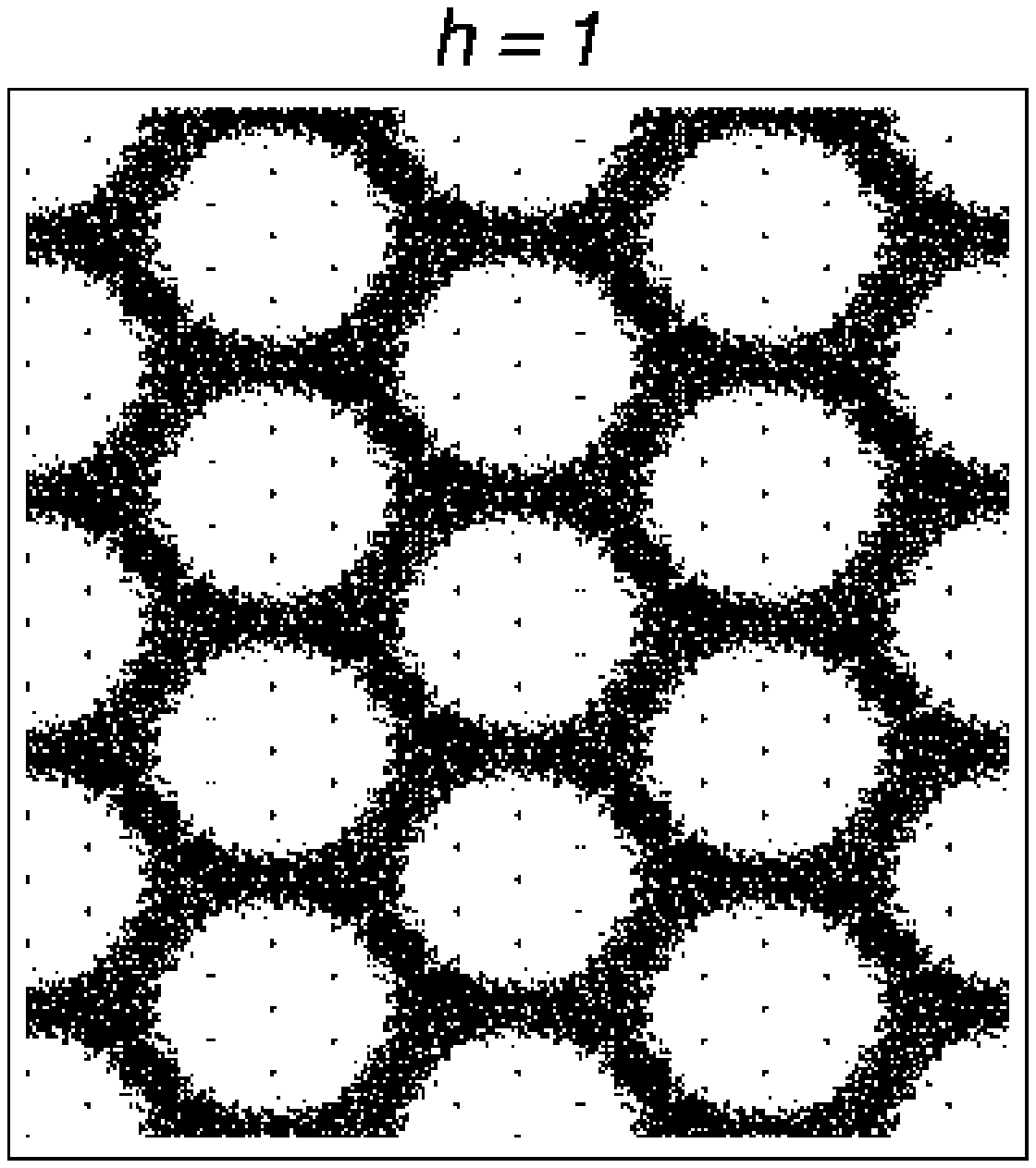}}
      \end{minipage}}
     \end{minipage}
  \end{center}
\caption{{\small Below the phase transition ($\rho=0.77$)}}
\label{fig:below}
\end{figure}

\begin{figure}[htb]
  \begin{center}
   \begin{minipage}{6cm}
     \epsfig{file=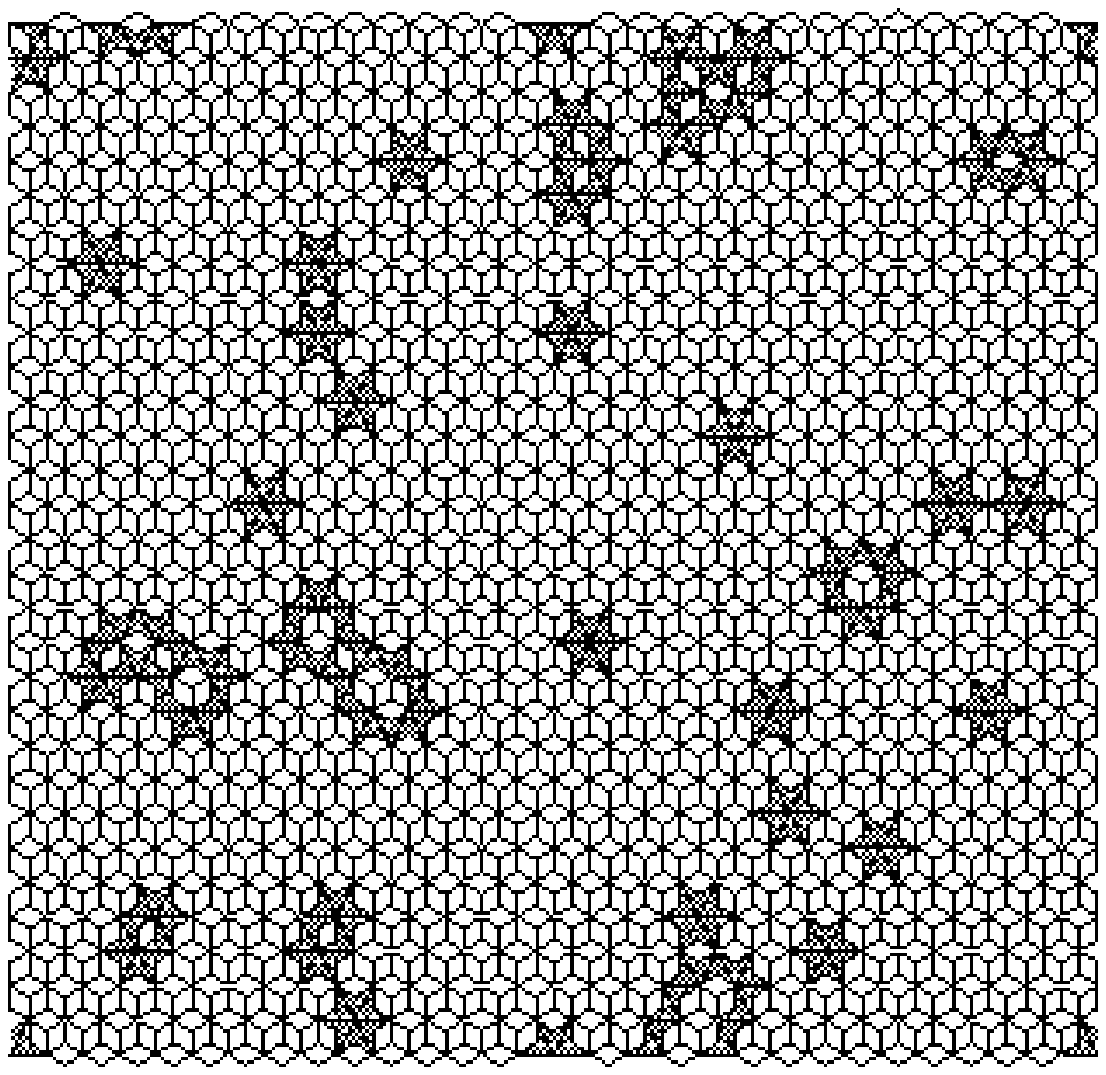,height=6cm}
   \end{minipage}
   \hfill
   \begin{minipage}{9cm}
    \centerline{
      \begin{minipage}{2.9cm}
        \centerline{\epsfysize=2.9cm \epsfbox{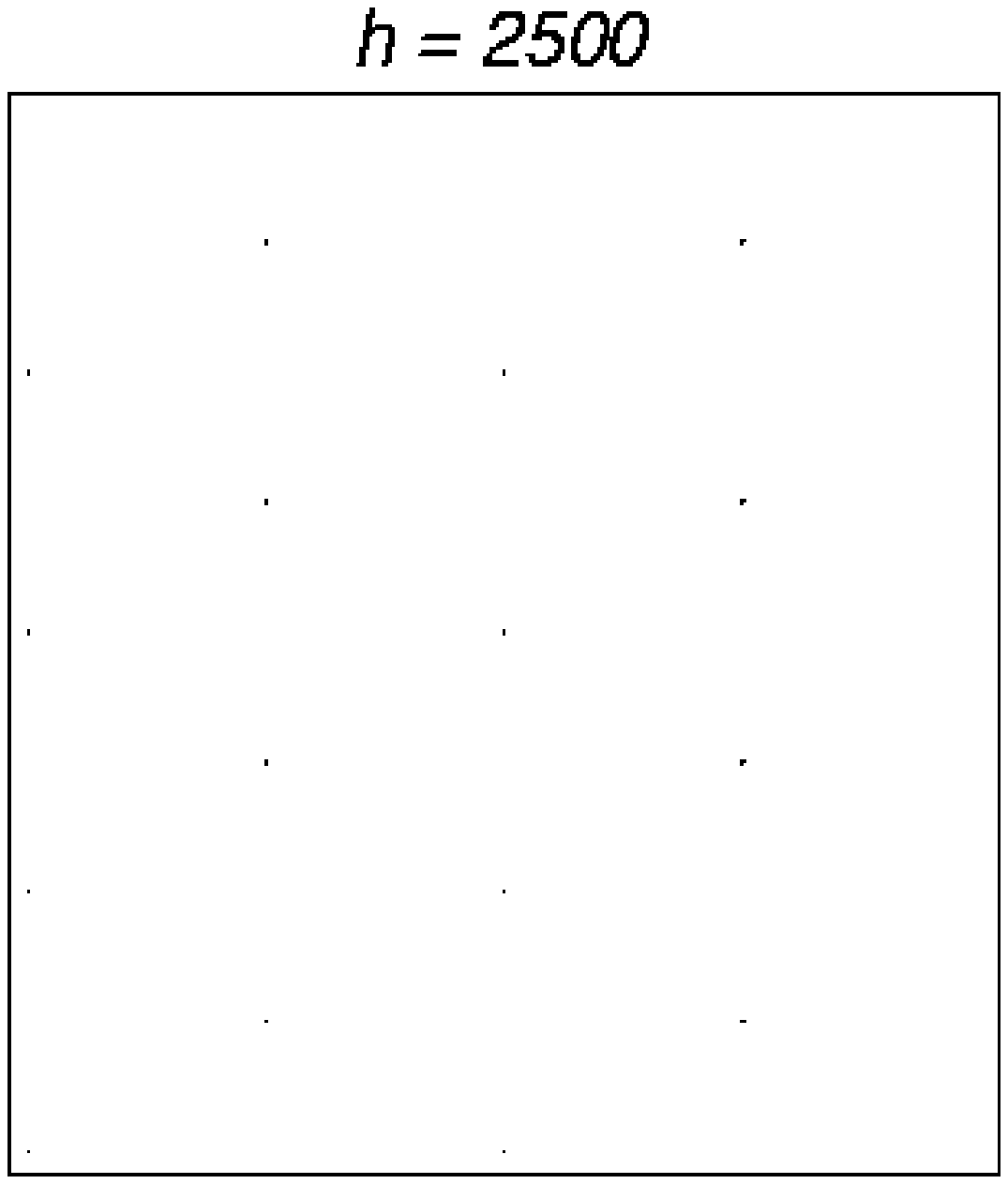}}
      \end{minipage}
      \hspace{1mm}
      \begin{minipage}{2.9cm}
        \centerline{\epsfysize=2.9cm \epsfbox{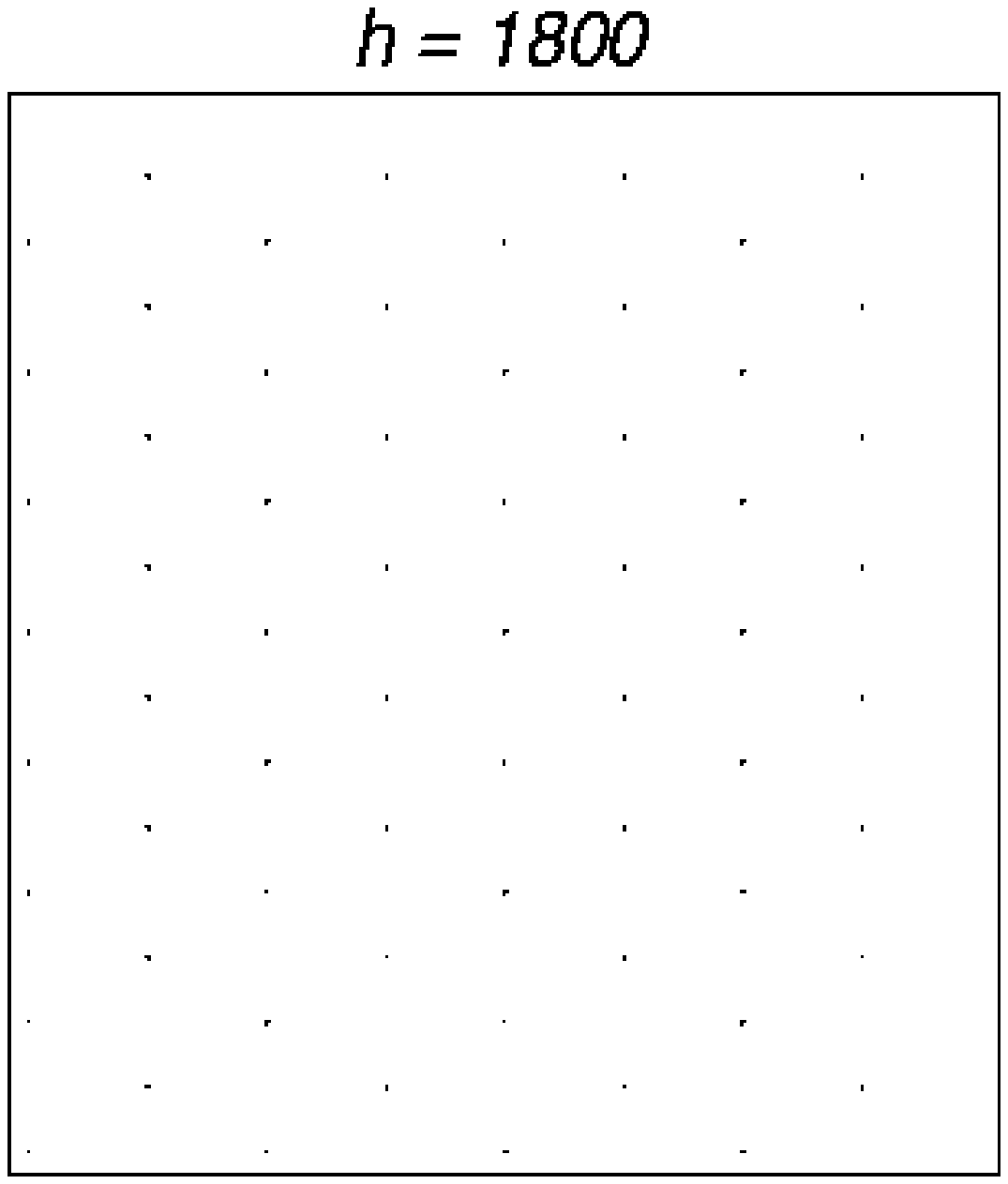}}
      \end{minipage}
      \hspace{1mm}
      \begin{minipage}{2.9cm}
        \centerline{\epsfysize=2.9cm \epsfbox{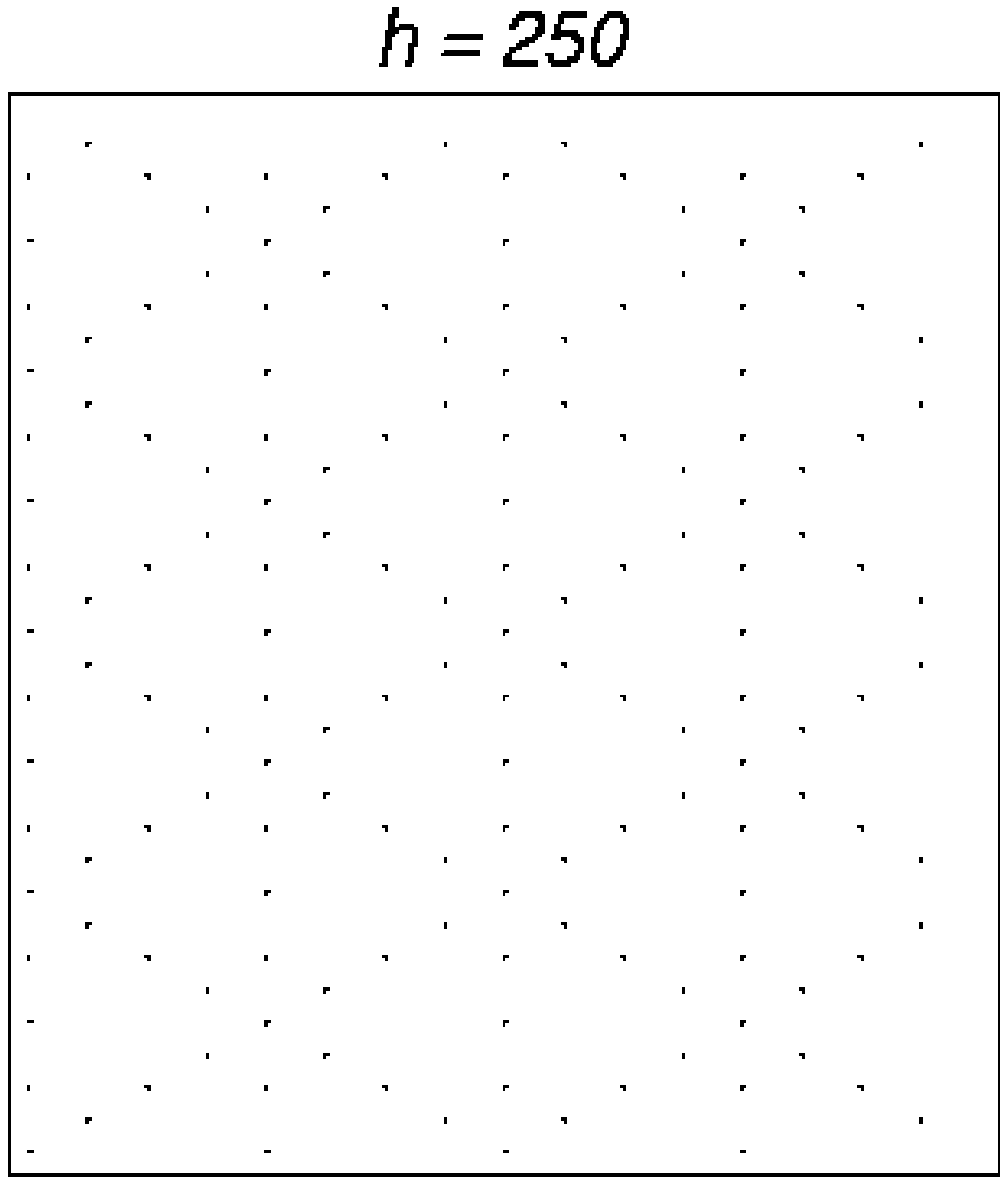}}
      \end{minipage}}
     \centerline{
      \begin{minipage}{2.9cm}
        \centerline{\epsfysize=2.9cm \epsfbox{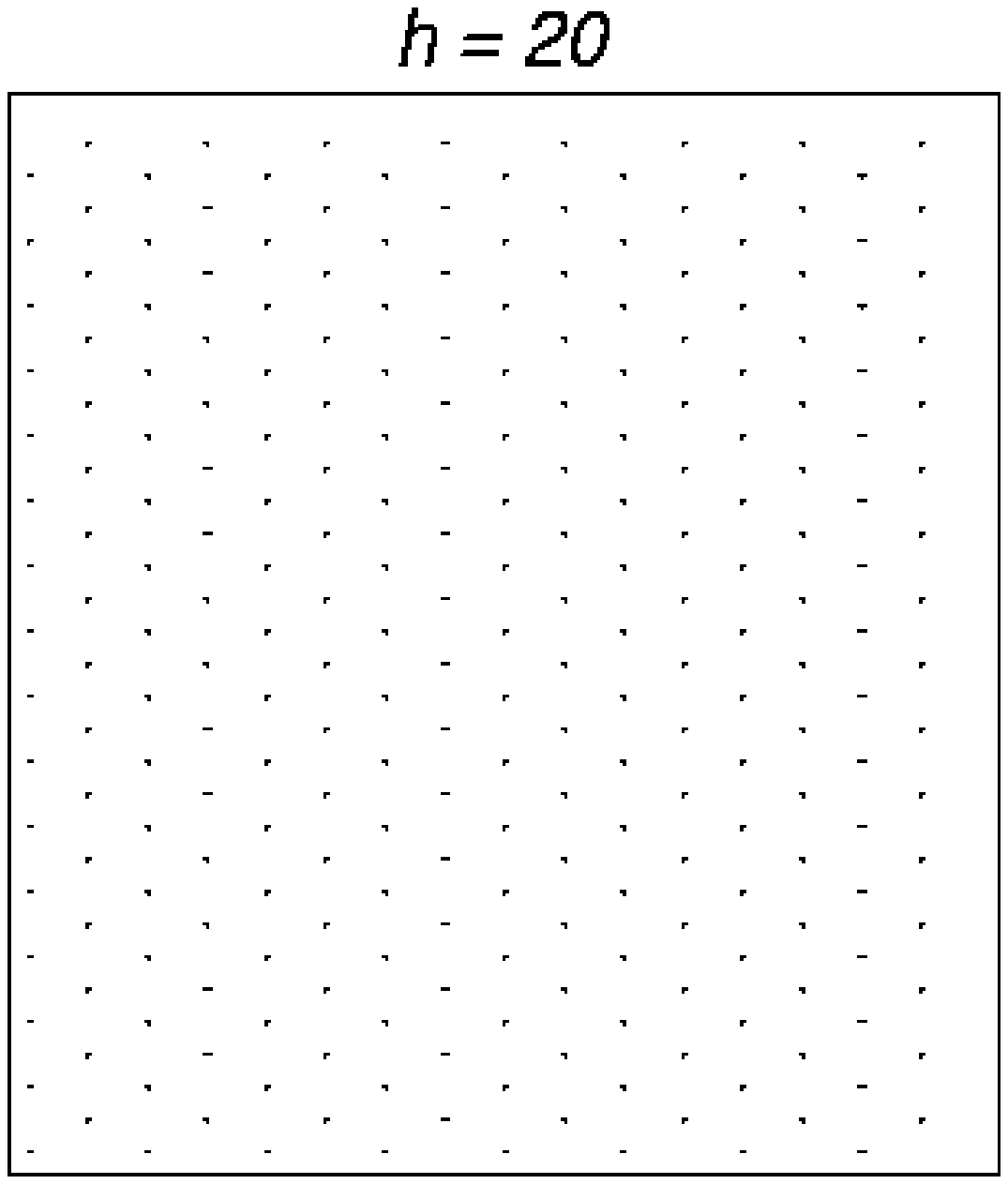}}
      \end{minipage}
      \hspace{1mm}
      \begin{minipage}{2.9cm}
%        \centerline{\epsfysize=2.9cm \epsfbox{pic3/Fourier34.eps}}
        \centerline{\epsfysize=2.9cm \epsfbox{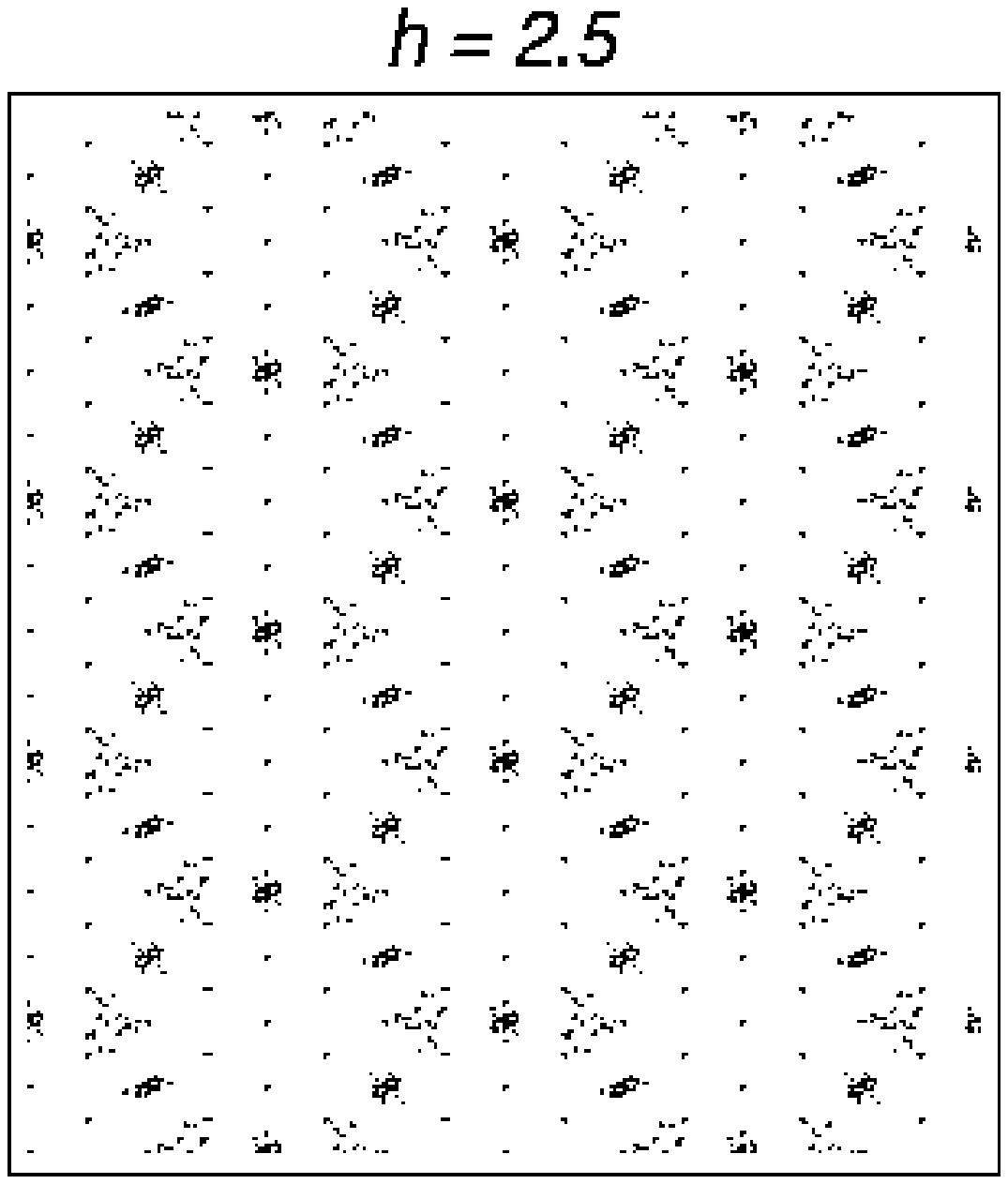}}
      \end{minipage}
      \hspace{1mm}
      \begin{minipage}{2.9cm}
%        \centerline{\epsfysize=2.9cm \epsfbox{pic3/Fourier34.eps}}
        \centerline{\epsfysize=2.9cm \epsfbox{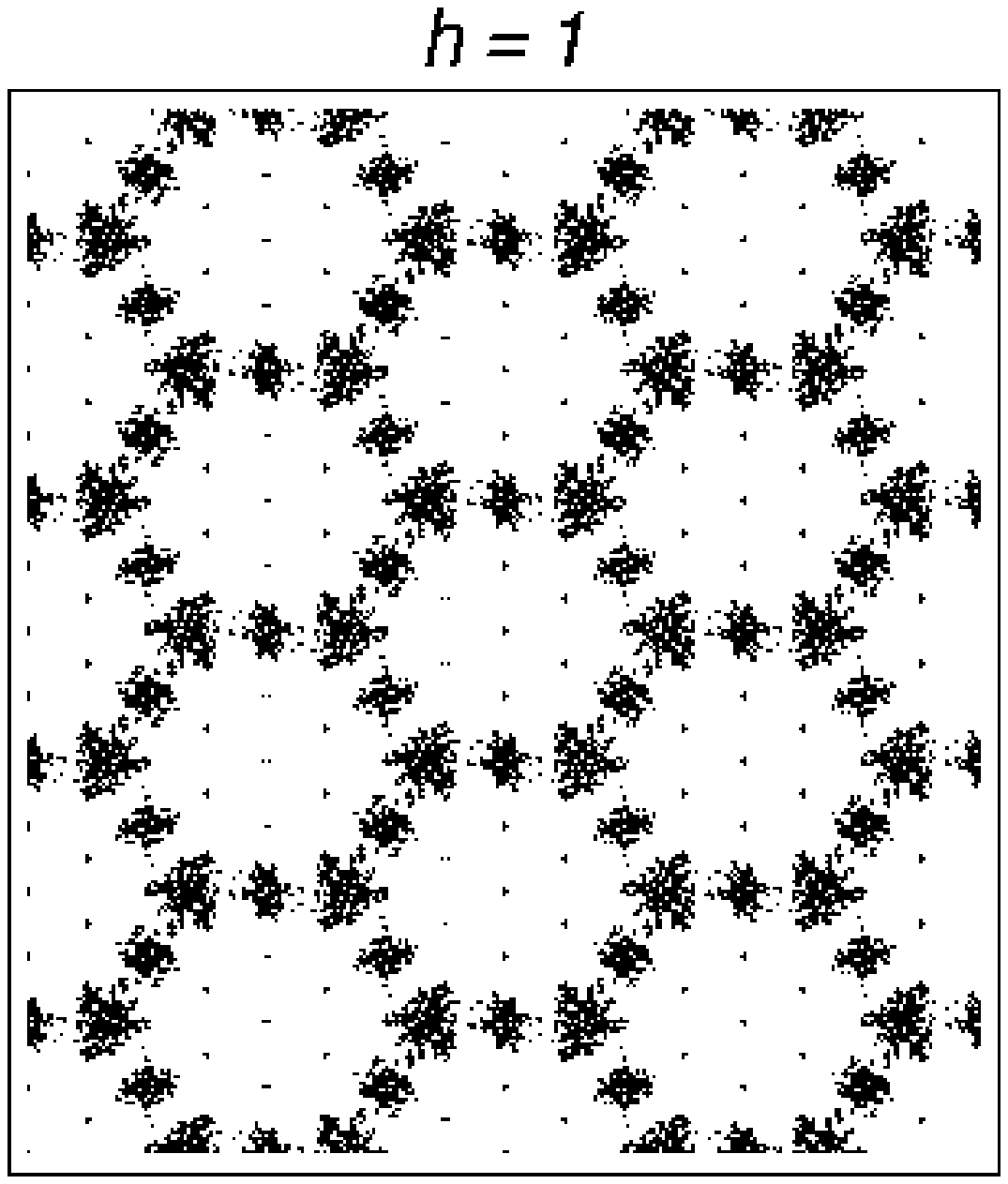}}
      \end{minipage}}
     \end{minipage}
  \end{center}
\caption{{\small Above the phase transition ($\rho=0.90$)}}
\label{fig:above}
\end{figure}

We now present another random tiling of the domino class, but this time with two 
different types of prototiles \cite{BHHR}.
It is therefore slightly more interesting than the model discussed above.
The prototiles of the dart-rhombus random tiling are $60^{\circ}$-rhombi and darts made from
two rhombus halfs, see figure \ref{fig:darttile}. 
In addition to the usual face-to-face condition, we impose an alternation condition
on the rhombi, such that neighbouring rhombi of equal orientation are excluded.
Finally, to avoid pathologic lines of alternating darts, we demand that two
neighbouring darts must not share a short edge.
These tiling rules force the darts to arrange in closed lines, see figures
\ref{fig:below} and \ref{fig:above}.
We thus deal with a {\em dilute loop model} of dart loops in a background of rhombi.
For more on this picture, which we will not expand here, see \cite{Ric}.
The dense loop phase has the minimal total rhombus density $\rho=\frac{1}{3}$, see 
also figure \ref{fig:minimum}.

This random tiling corresponds to the fully packed dimer system on one of the so-called
Archimedian tilings (where all faces are regular polygons) \cite{GS}.
Lattices and tilings of this type have already been described in detail by Kepler 
\cite{Kep}, but have since been rediscovered many times.
In the context of statistical mechanics the graph of our model has also been called
a Fisher lattice \cite{T}, its dual graph is named a Kagom\'e lattice.
The first solution of the Fisher lattice dimer model was given by Fan and Wu \cite{FW} 
by realizing the correspondence to a soluble subcase of the
the eight-vertex model \cite{B}, the so-called free-fermion case.
A one-to-one mapping between the free-fermion case of the eight-vertex model
\footnote{
We follow the Baxter's convention here \cite{B}.} 
and the dart-rhombus tiling is obtained by a reformulation
of the latter in terms of `supertiles', see figure \ref{fig:supertiles}.
\Bild{4}{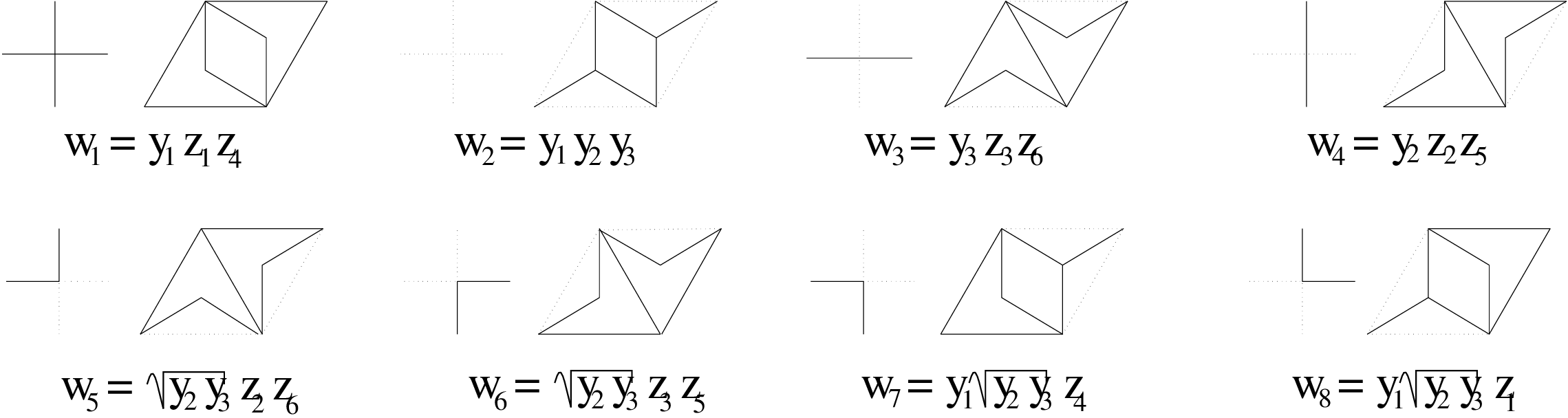}
{A mapping between the eight-vertex model and 
the dart-rhombus tiling via ``supertiles''.}
{fig:supertiles}
Closed dart loops in the tiling correspond to closed edge loops on the square lattice.

Another interpretation exists as a model with domain walls where different walls can meet
and annihilate \cite{Bha84}, in generalization of the dimer system on the hexagonal 
lattice, which can also be interpreted as a domain-wall model.
In our model, the domain walls correspond to the dart loops.

Let us denote the densities of rhombi by $\rho_i$ $(i=1,2,3)$ and the densities of
darts by $\sigma_i$ $(i=1,\ldots,6)$. 
We have several constraints on these macroscopic variables:
in addition to the filling condition, the loop condition means that densities of 
opposite darts have to be the same,
\be 
\sigma_1=\sigma_4, \qquad \sigma_2=\sigma_5, \qquad \sigma_3=\sigma_6. 
\ee
Moreover, as each dart is accompanied by a rhombus of equal type, the remaining rhombi
occur with equal frequency of each type due to the alternation condition,
\be 
\rho_1-\sigma_1=\rho_2-\sigma_2=\rho_3-\sigma_3. 
\ee
These constraints reduce the number of independent parameters to three.

The group of geometric symmetries of the tiling is $D_3$.
But these are not the only symmetries of the model: 
there is a hidden symmetry which exchanges darts and rhombi.
It is most easily described in the vertex model formulation where it corresponds to
the exchange of free and occupied edges (spin reversal symmetry).

The full symmetry group of the dart-rhombus tiling turns out to be the 
{\em tetrahedral group} $T_d$.
As the group action respects the density constraints and acts irreducible on the
remaining subspace of independent density parameters, we can characterize the phase
space of the dart-rhombus random tiling according to the three-dimensional irreducible
representation of the tetrahedral group $T_d$.
Each phase of the dart-rhombus tiling corresponds to a point inside the tetrahedron:
we fix the tetrahedron center as the origin and choose an orthonormal basis of unit vectors
pointing to edges along twofold axes, as is indicated in figure \ref{fig:tetraeder}.
\Bild{6}{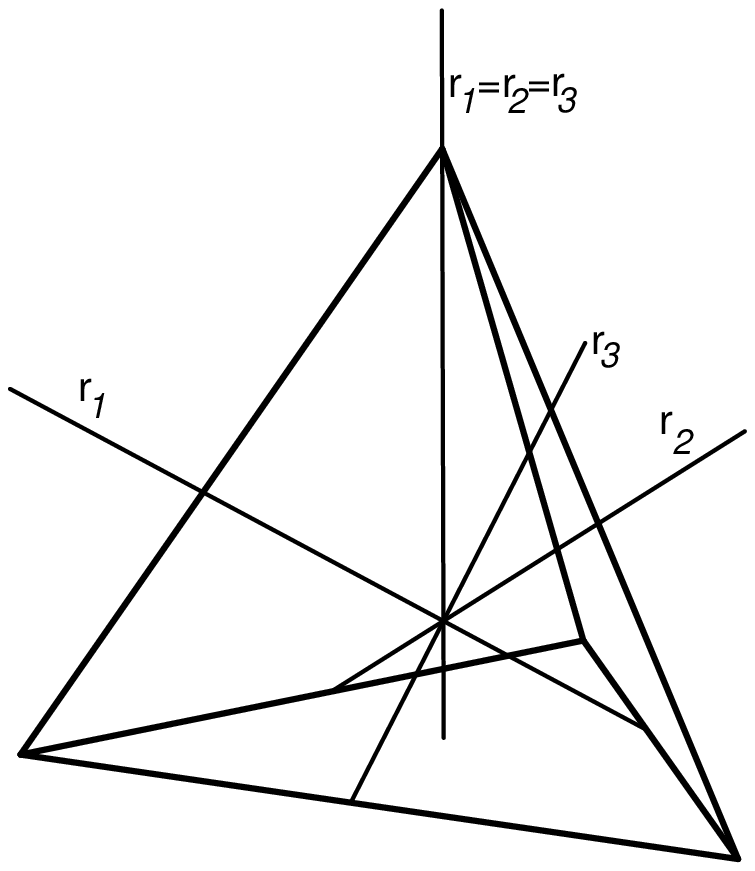}{The phase space of the dart-rhombus tiling is a tetrahedron.}{fig:tetraeder}
In this way we obtain a parametrization in terms of reduced rhombi densities
\be 
r_i=\rho_i-\frac{1}{6} \qquad (i=1,2,3). 
\ee
The vectors pointing to the four corners of the tetrahedron are unit vectors for the
reduced dart densities $s_i$ and for the reduced total rhombi density $r$,
\bea &s_i = 2\sigma_i-\frac{1}{6} \qquad (i=1,2,3),& \\
 &r=r_1+r_2+r_3.& \nonumber \eea

In this geometry, the dense loop phase is found on the triangle surface 
with minimal rhombus density $r=-\frac{1}{6}$.
The other triangle surfaces which are obtained by the exchange symmetry are diluted
phases where the darts arrange along directed lines due to exclusion of one 
dart orientation.
On the other hand, the dense loop phase is an obvious decoration of the 
{\em rhombus tiling}, suggested by figure \ref{fig:minimum}.
We conclude that the surface phases are of Kasteleyn type with phase transitions
at the corners of the tetrahedron.

The point of maximum entropy (the tetrahedron's centre) is fixed by symmetry to
$\rho_i=\frac{1}{6}$, $\sigma_i=\frac{1}{12}$ where the darts and the rhombi occupy
half of the tiling area each.
The value at the entropy at its maximum
\be 
s_{\rm max} = \frac{1}{3}\log 2 \approx 0.231 \label{form:drent}
\ee
follows from a geometrical consideration:
in the supertile formulation, building a rhombic patch by starting from a prescribed 
configuration on the left and on the upper boundary of a $\frac{\pi}{3}$ sector, 
there are in general two
different possibilities to add a new tile.
This results in an entropy of $\log 2$ per supertile because the choice of the
boundary does not affect the entropy value.
Observing that each supertile consists of 3 ordinary tiles by area, one obtains
(\ref{form:drent}).

In general, phase transitions occur in regions where we have infinite dart lines
which tend to minimize the enclosed area as in percolation, see figures 
\ref{fig:below} and \ref{fig:above}.
For a more concrete description, we introduce rhombus activities $y_1, y_2, y_3$, 
dart activities $z_1,\ldots, z_6$,
and compute the grand-canonical potential per tile via Pfaffians, yielding \cite{T}
\be 
\phi=\frac{1}{24\pi^2}\int\limits_0^{2 \pi}\int\limits_0^{2 \pi} \log 
\det \lambda (\varphi_1,\varphi_2) d\varphi_1 d\varphi_2, \label{form:dartrhombus}
\ee
where the determinant is given by
$$
\begin{array}{lrcl} \hspace*{-3mm}&
\det \lambda (\varphi_1,\varphi_2)&=& y_1^2 z_1^2 z_4^2 + y_2^2 z_2^2 z_5^2
+ y_3^2 z_3^2 z_6^2 + y_1^2 y_2^2y_3^2 \\
& & &+ 2 y_1 y_2(z_1 z_2 z_4 z_5 -y_3^2 z_3 z_6) \cos\varphi_1 \\
& & &+ 2 y_2 y_3(z_2 z_3 z_5 z_6 -y_1^2 z_1 z_4) \cos(\varphi_1-\varphi_2) \\
& & &+ 2 y_3 y_1(z_1 z_3 z_4 z_6 -y_2^2 z_2 z_5) \cos\varphi_2 .
\end{array} 
$$ 
This result was obtained by using {\em periodic} boundary conditions.
It applies, however, in the case of nonzero activities, to the free case as well, 
since this has been shown by Ruelle \cite{R} for the more general eight-vertex model.
(Note that the Kasteleyn subclass has already been discussed above.)

Let us now proceed with the discussion of the critical behaviour as a function
of the tile densities.
It is of Onsager type in the generic case \cite{FW}.
At points fulfilling the criticality condition 
\be w_1+ w_2+ w_3+ w_4=2\max\{w_1,w_2,w_3, w_4\} \ee
the grand-canonical potential is nonanalytic, resulting in a logarithmic divergence in
the covariance matrix.
The $w_i$ $(i=1,\ldots,8)$ are the Boltzmann factors of the eight-vertex model whose
relation to the tile activities is given in figure \ref{fig:supertiles}.

The entropy as a function of the rhombi densities indeed leads to a Hessian which
is proportional to the identity.
In the four-dimensional vector space of the reduced dart densities and the reduced
total rhombus density, the entropy is defined on the hyperplane complementary to the
vector $(1,1,1,1)^t$. 
This leads to
\be I(r,r) = 2(r_1^2+r_2^2+r_3^2), \qquad \lambda = 6.\ee

Let us focus on the behaviour of the random tiling in the sixfold symmetric phase
where each orientation of darts, respectively rhombi, occurs with the same frequency.
The only free parameters are the total density $\rho$ of the rhombi or the
total density $\sigma$ of the darts, each being equally distributed over the different
orientations.
The entropy curve as a function of the rhombus density $\rho$ is plotted in figure 
\ref{fig:dartent}.
\Bild{6}{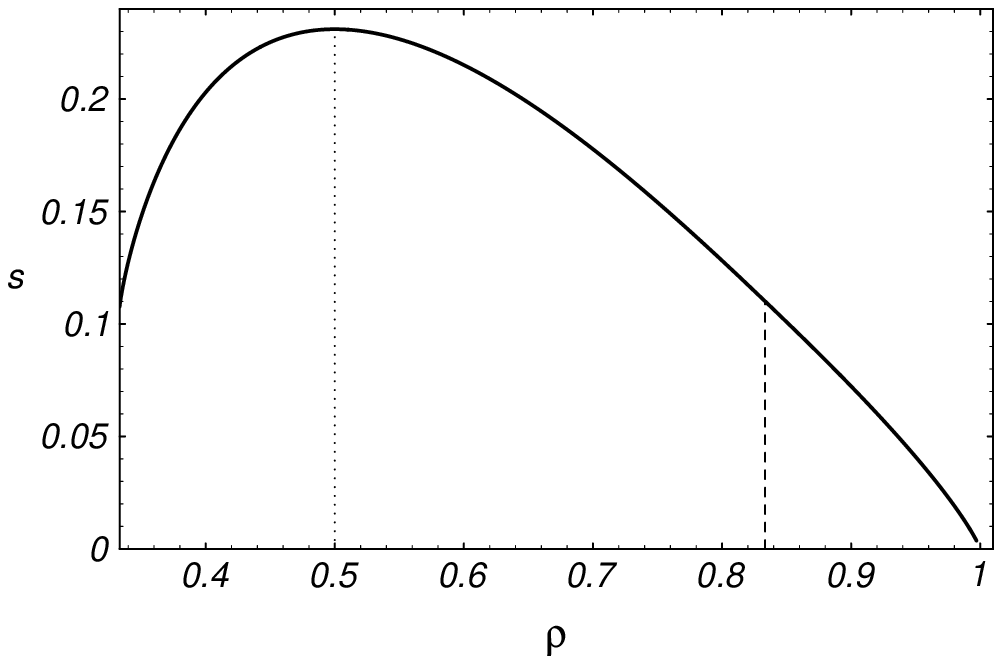}{Entropy of the dart-rhombus random tiling in the sixfold symmetric
phase as a function of the rhombus density $\rho$.}{fig:dartent}
The maximum is already discussed above; 
the elastic constant in the sixfold phase is $\lambda=6$.
A phase transition of Onsager type occurs at $\rho=\frac{5}{6}$ with the usual
logarithmic singularity in the isothermal compressibility.
A phase transition of Kasteleyn type occurs at $\rho=1$.
The point $\rho=\frac{1}{3}$ is the point of maximal symmetry of the previously
mentioned rhombus tiling (its maximum scaled by a factor of 3).

\begin{figure}[htb]
  \begin{center}
   \hspace{-1cm}
   \begin{minipage}{6cm}
     \epsfig{file=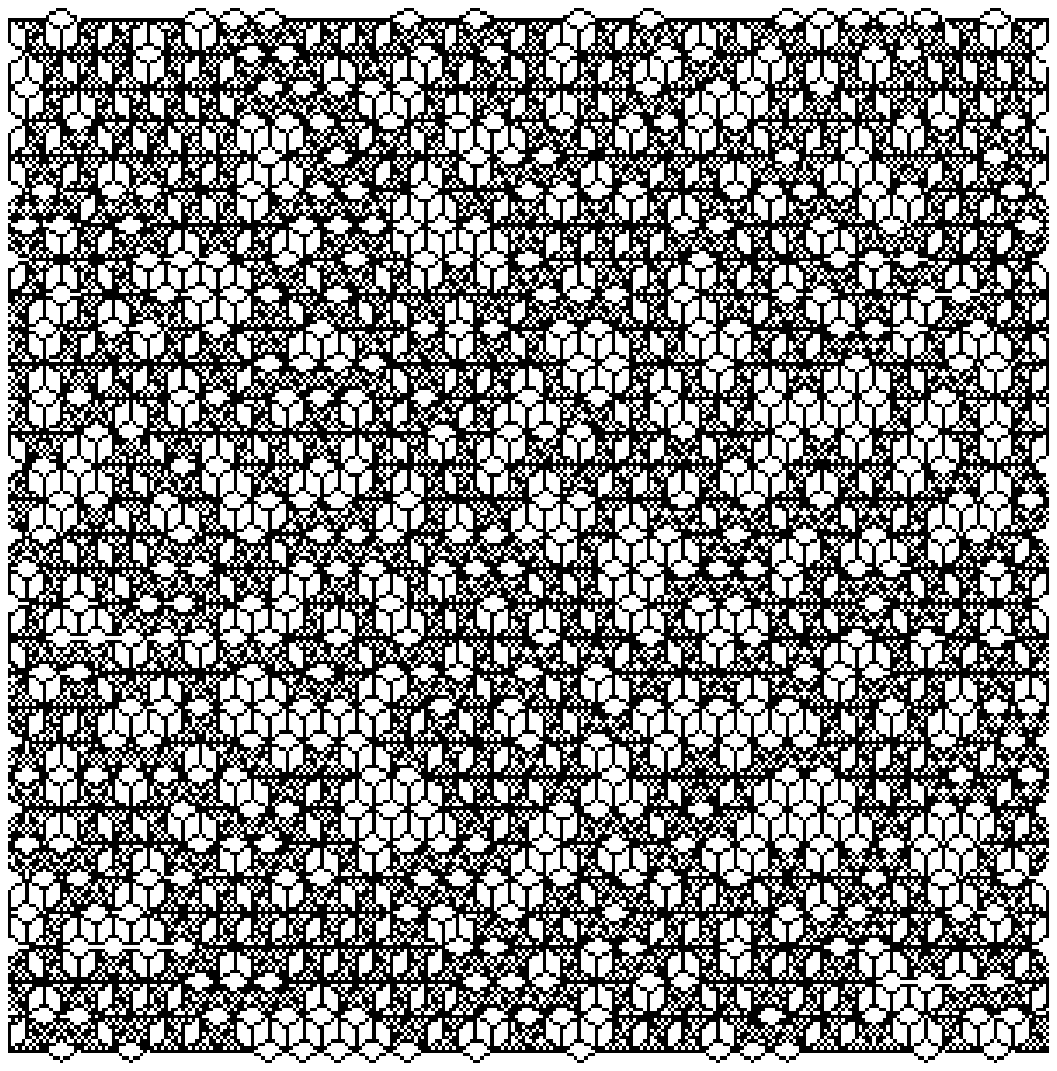,height=6cm}
   \end{minipage}
   \hspace{1cm}
   \begin{minipage}{5.9cm}
    \centerline{
      \begin{minipage}{2.9cm}
        \centerline{\epsfysize=2.9cm \epsfbox{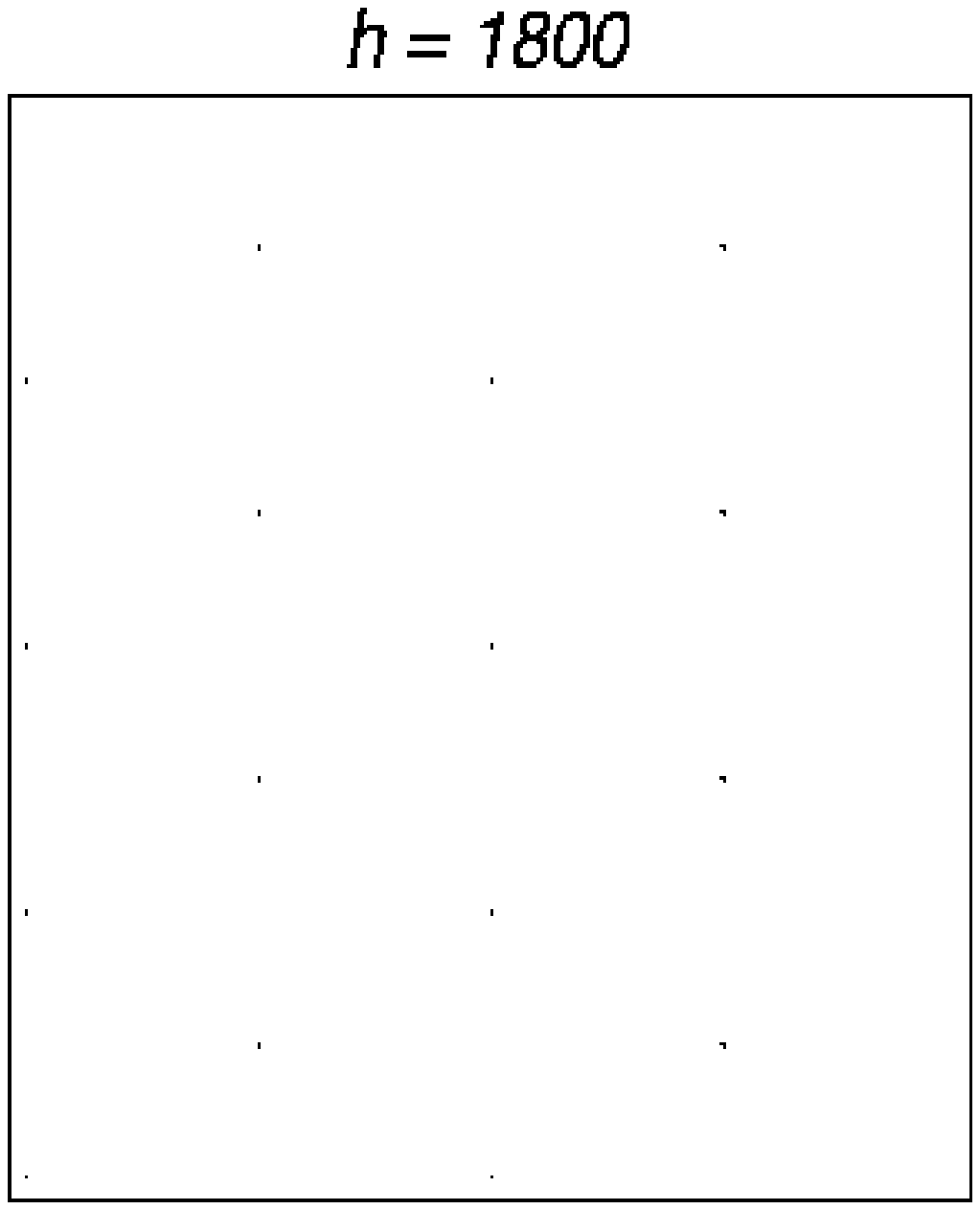}}
      \end{minipage}
      \begin{minipage}{2.9cm}
        \centerline{\epsfysize=2.9cm \epsfbox{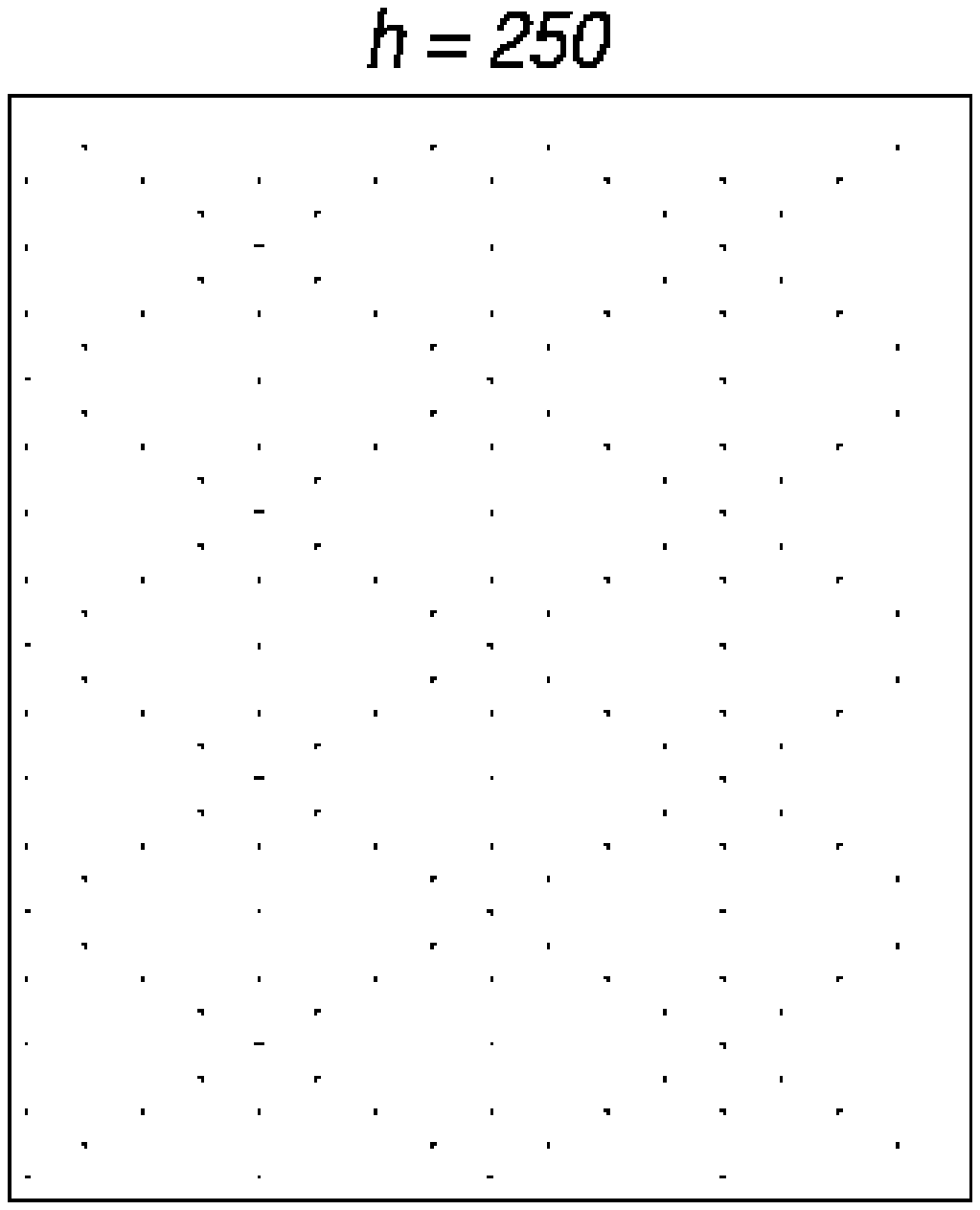}}
      \end{minipage}}
     \centerline{
      \begin{minipage}{2.9cm}
        \centerline{\epsfysize=2.9cm \epsfbox{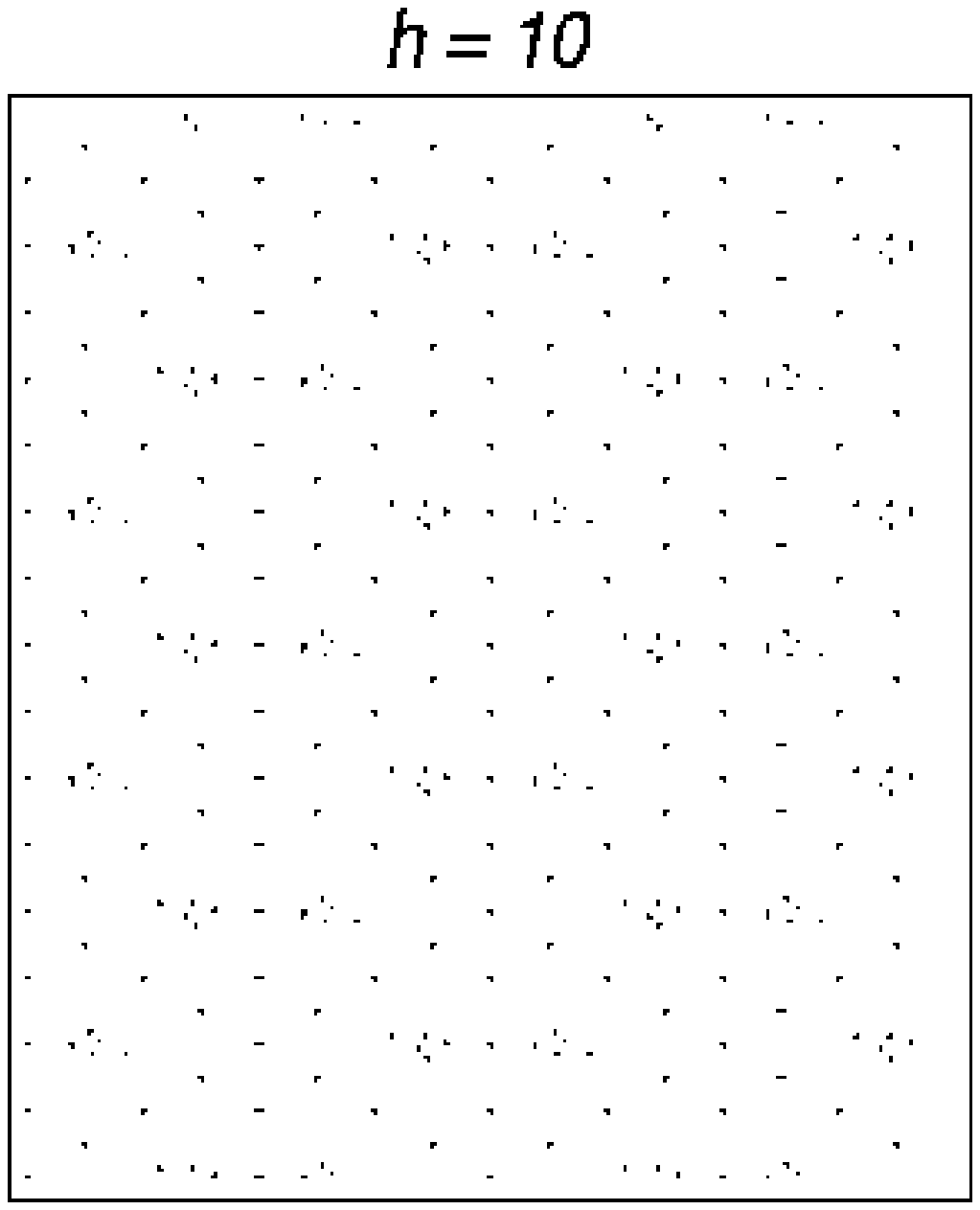}}
      \end{minipage}
      \begin{minipage}{2.9cm}
%        \centerline{\epsfysize=2.9cm \epsfbox{pic4/Fourier43.eps}}
        \centerline{\epsfysize=2.9cm \epsfbox{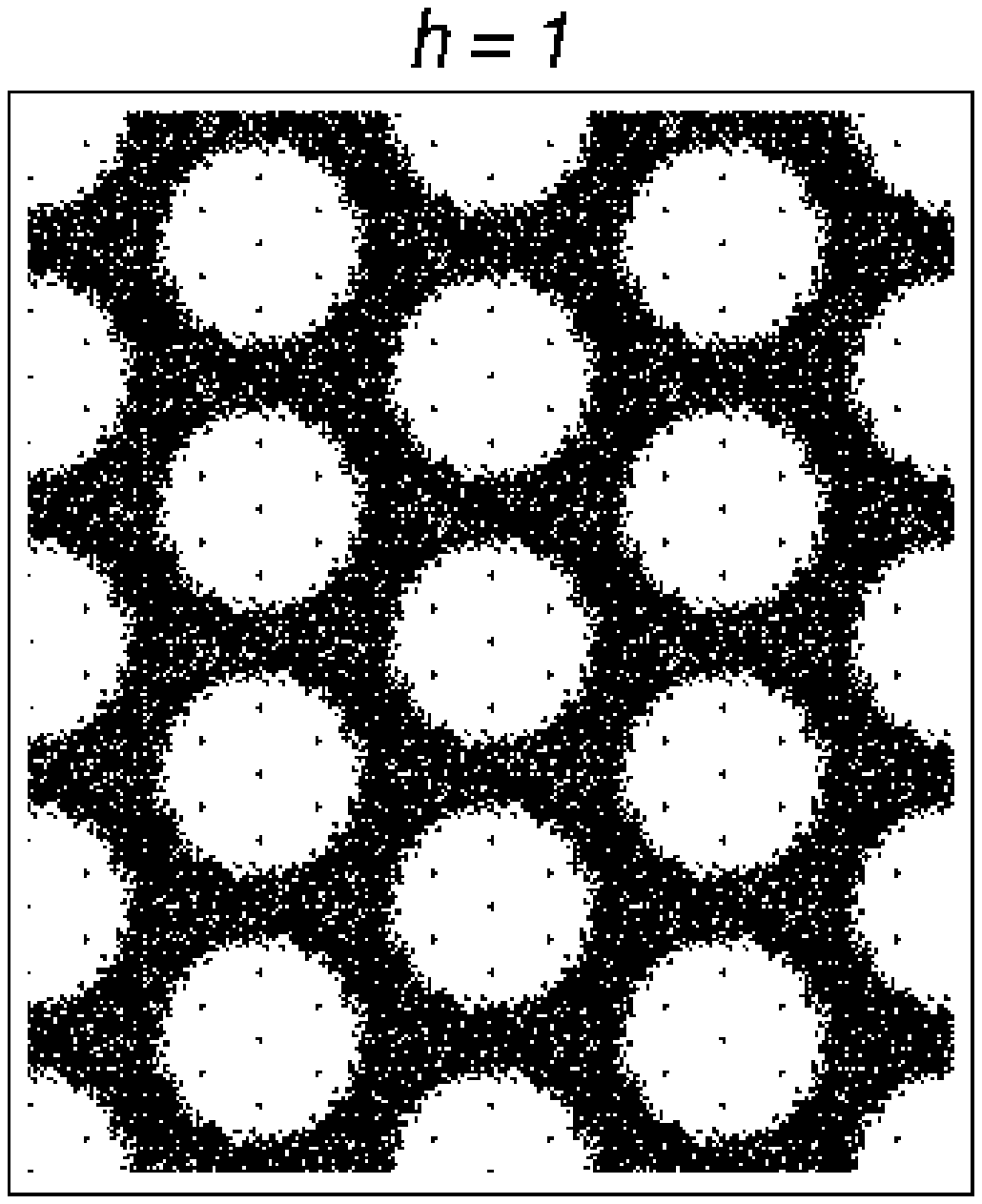}}
      \end{minipage}}
     \end{minipage}
  \end{center}
\caption{{\small The entropy maximum ($\rho=0.50$)}}
\label{fig:maximum}
\end{figure}

Below are four characteristic snapshots of tilings in the
sixfold symmetric phase together with their diffraction patterns, taken
at different rhombi densities \cite{HO}.
The tilings are periodic in the vertical direction and periodic up to a cyclical shift
in the horizontal direction.
To obtain diffraction patterns, delta-scatterers were positioned in the middle of every
prototile.
Figures \ref{fig:minimum}-\ref{fig:above} show contour lines of the absolute value of the 
scatterer's Fourier transform, computed in arbitrary units.

The phase transition (at rhombi density $\rho_c=\frac{5}{6}$) is directly visible 
in the tiling.
Above the phase transition ($\rho>\rho_c$) we have just some small
dart islands (figure \ref{fig:above}) whereas below the phase transition ($\rho<\rho_c$)
they form infinite lines as in 
percolation (figure \ref{fig:below}). 
The phase transition is also visible in the diffraction pictures.
Above the phase transition point, we have a small separated, sixfold symmetric
background which gets connected below the phase transition. 

At the entropy maximum (figure \ref{fig:maximum}), the diffuse background has maximal
intensity and is least structured whereas at minimal rhombus density 
(figure \ref{fig:minimum}) it is still connected but laced up. 

We mention that the height of the diffraction spots scales with the system size, as 
follows from our numerical analysis.
The diffraction pattern in the thermodynamic limit consists of a point part 
(Bragg peaks) and an absolutely continous background \cite{BH98}.
Although this is not the generic result one expects for 2D random tilings, it is 
inevitable for this kind of crystallographic random tiling that lives on a lattice
background.

\subsubsection{Three-colourings of the square lattice -- a monomino tiling violating the
second random tiling hypothesis}

The three-colouring model is the ensemble of tilings with
three types of monominoes such that no two tiles of the same type are adjacent.
Since the packing rule is of nearest-neighbour type, the restriction is hierarchical,
and it is instructive to look at the one-dimensional model first.
In one dimension, using the recursion method demonstrated  above, the grand-canonical 
potential can be shown to be the logarithm of the positive real root of the equation
\be x^3-(z_1z_2+z_2z_3+z_3z_1)x-2z_1z_2z_3=0, \ee
the $z_i$ being the activities of the different monominoes.
By construction, the entropy of this ensemble is an upper bound for the
entropy of its generalization to two dimensions.

The group $S_3$ of colour permutations is a symmetry for both models which 
also respects the filling constraint $\rho_1+\rho_2+\rho_3=1$.
Consequently, the maximal entropy occurs at $\rho_1=\rho_2=\rho_3=\frac{1}{3}$, 
and the thermodynamic functions are invariant under the two-dimensional irreducible 
action of the symmetry group in the subspace of independent densities.
In particular, there is only one single elastic constant, which is $\lambda = 9$ for 
the one-dimensional model.

Bounds for the maximal entropy can be obtained as follows.
The configurations with highest frequency of monominoes of type $a$ are arrangements 
where all $a$'s fill one sublattice.
As the other vertices can independently be filled by monominoes of type $b$ or $c$, 
we get a lower bound for the entropy: $s_{\rm max} \ge \log\sqrt 2$.
The one dimensional problem, at its point of maximal entropy, is a genuine 
Bernoulli type ensemble of entropy $\log 2$, so we found
\be 
0.3466\approx\log\sqrt2 \le s_{\rm max} \le \log 2\approx 0.6931. 
\ee

The given counting problem has first been solved by Baxter 
\cite{B1} using Bethe's Ansatz. 
Introducing chemical potentials $\mu_1, \mu_2, \mu_3$ for the different monominoes, 
the grand-canonical potential reads \cite{B}
\be 
\phi(\mu_1, \mu_2, \mu_3)  = 
\log \frac{ 8 (z_1 z_2 z_3)^{\frac{1}{3}}}{ \sqrt{\frac{(u-w)^3}{v}} - 
\sqrt{\frac{(v-w)^3}{u}} }, 
\ee
where $u,v,w$ are the positive real solutions of
\be 
x(3B-x)^2=4, \qquad u > v \ge w,
\ee
where $B$ is given by
\be 
B = \frac{z_1 z_2 + z_2 z_3 + z_1 z_3}{3 (z_1 z_2 z_3)^{\frac{1}{3}}}.
\ee
It has to be stressed again that this solution was obtained by making use of 
{\em periodic} boundary conditions.
The result agrees, on the other hand, with the corresponding free limit, as can
be shown \cite{Ric} by generalizing an argument of G. Kuperberg \cite{Kup97}.

The value at the maximum entropy is
\be 
s_{\rm max}=\frac{3}{2}\log \frac{4}{3} \approx 0.4315, 
\ee
which was already obtained before as the residual entropy of square ice \cite{L}.

It is instructive to analyse the entropy function along a line of reflection symmetry,
 see figure \ref{fig:colent}.
\Bild{6}{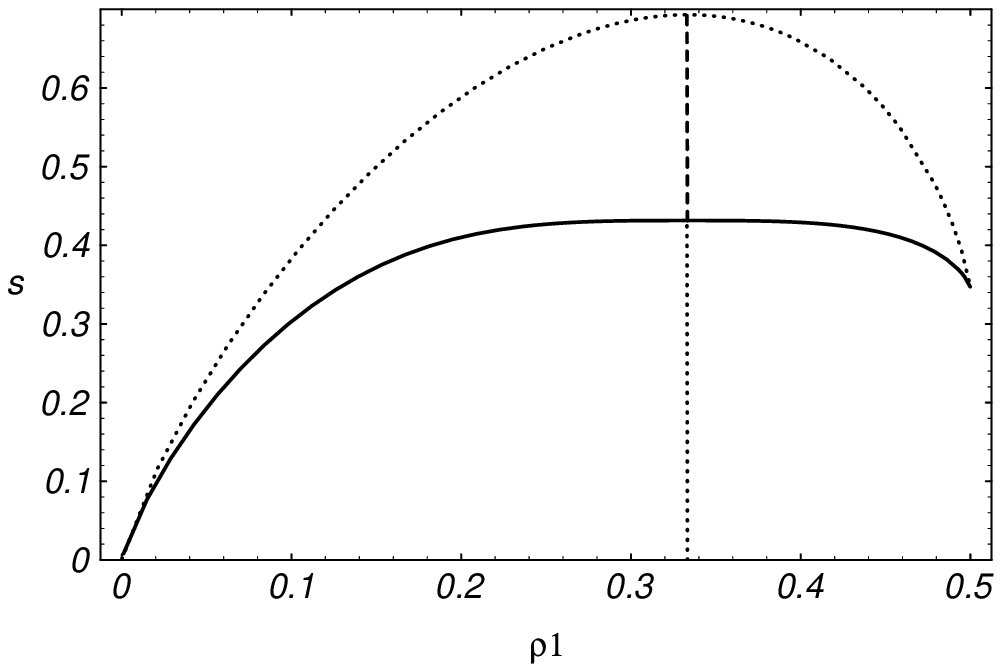}
{Entropy of the three-colouring model along a line of reflection symmetry
($\rho_2=\rho_3$) as a function of $\rho_1$ (heavy curve).
The dotted curve represents the system with exclusion restricted to one dimension.}
{fig:colent}
Note that the entropy at its maximum is flat, indicating a vanishing second derivative, 
hence a phase transition at $s=s_{\rm max}$.
The phase transition is of fluid-solid type: 
below the phase transition, monominoes arrange on the lattice homogeneously, whereas 
above the phase transition one sublattice is preferred in occupation \cite{PS89}.
The phase transition causes a square-root singularity in the covariance
matrix \cite{B1}.

In fact, explicit expressions for the entropy can be given in the two phases \cite{Ric}.
It turns out that both functions are analytic on the whole interval $0 \le \rho < \frac{1}{2}$
and intersect smoothly at $\rho_1=\frac{1}{3}$.
They can therefore be polynomially approximated at the entropy maximum.
The absolute values of the coefficients coincide up to fourth order;
with $r=\rho-\frac{1}{3}$, we find
\be s(r) = \frac{3}{2}\log \frac{4}{3} -\frac{1}{3!} \frac{9^2}{2} |r|^3 -
\frac{1}{4!} \frac{9^3}{2} r^4 + {\cal O}(r^5).
\ee
We conclude that the square lattice three-colouring model violates the second random 
tiling hypothesis!
This means that the usual reasoning about elastic theory \cite{Hen91} is not applicable here.

It would be interesting to compare this result with an expansion of the entropy 
in terms of variables taken from the height representation of this model \cite{BH97}.

\subsubsection{Hard hexagons -- a monomino tiling with unexpected entropy}

Our last example belongs to a tiling class with two kinds of monominoes which we call
particles and holes together with nearest neighbour exclusion for the particles.
Its one-dimensional version has already been discussed above.
The two-dimensional generalizations to the square lattice and to the hexagonal 
lattice correspond to the ensembles of {\em hard squares} \cite{BET, GF} and 
{\em hard hexagons} \cite{B2,B}.
Both models show a kind of fluid-solid phase transition:
the low density phase is homogeneous, each sublattice has the same particle density,
whereas at high density one sublattice is preferred in occupation. 
There is no obvious symmetry in these models, and therefore no preferred point
in the phase space where the entropy maximum should be located.

It is interesting to mention that the three-colouring model discussed above was 
interpreted as an `approximating' hard squares model, with two types of holes 
\cite{B1, PS89}.
As the hard squares model itself has escaped exact solution so far, we concentrate 
on the hard hexagon model in the following.
The solution of this model was given by Baxter \cite{B2,B}.
This result is again obtained from {\em periodic} boundary conditions. 
It is easy to see, however, that it coincides with the result for the 
corresponding free boundary conditions, 
as each free rectangular patch can be transformed into a periodic one by adding a 
row of holes.

A phase transition occurs at $\rho_c=\frac{1}{\tau\sqrt5}\approx 0.2764$ 
(the maximal density being normalized 
to $\frac{1}{3}$) with a critical exponent $\alpha=\frac{1}{3}$ in the covariance matrix.
The entropy is a concave function with a quadratic maximum
\be s_{\rm max}\approx 0.333243 \quad \mbox{at} \quad \rho\approx 0.162433. \ee
Note that the maximum occurs nearly at the point where the hexagons occupy half
of the lattice area.
The small deviation of the entropy from $\frac{1}{3}$ lead to an incorrect 
prediction
\footnote{
For a short history of the model, see chapter 14.1 of Baxter's book \cite{B}.
} in 1978.
The elastic constant turns out to be $\lambda\approx 24.0807$.

Joyce \cite{J} used the theory of modular functions to write the particle 
activity $z$ as an {\em algebraic} function of the mean density $\rho$.
Applying Joyce's results, one can in fact find closed expressions for $s_{\rm
max}, \rho$ and $\lambda$.
As these are quite complicated, we omit them here \cite{Ric}.

\subsection{Concluding remarks}

We have shown, both conceptually and by a number of examples, that the densities of
tiles in random tiling ensembles provide natural parameters to investigate the phase
diagram of such models.
This is then independent of the existence of a height representation, but fully
consistent with it if it happens to exist.

Whereas the re-interpretation of exactly solvable models directly leads to a number
of random tilings with analytic expressions for the entropy function, it is much more
difficult to extract information about diffraction properties via Fourier transforms
because this also requires the knowledge of the autocorrelation.
For simple models such as the domino tiling and the rhombus tiling, it is possible
\cite{BH98} to compute the diffraction Bragg part explicitly and also the
behaviour of the diffuse background in the thermodynamic limit.

The next step in the analysis of random tilings is certainly the application of
the concept developed so far to quasicrystalline random tilings.
Although there are recent exact results on rectangle-triangle tilings 
\cite{Gie96, Kal94}, the interpretation of some of the results, for
example the shape of the entropy curve in the solvable regime, is subject to
controversial discussion.
It may turn out that the framework of statistical mechanics proposed here gains
additional insight into the phase space of the tilings.

\subsection*{Acknowledgements}

It is our pleasure to thank Paul Pearce for several useful hints and some help with
the body of known results and Elliott Lieb for a number of clarifying discussions.
We thank Chris Henley for numerous valuable comments on the manuscript.
Financial support from the German Science Foundation (DFG) is gratefully acknowledged.

\subsection*{Appendix: Height representation of the rhombus tiling}
Here, we briefly describe the connection between our description of the rhombus tiling to the 
description by Henley \cite{Hen88,Hen91}.
In particular, we show the connection between the elastic constants defined by both
approaches.

As was first pointed out in \cite{BH}, each configuration of the rhombus tiling can 
be obtained by projecting a roof-type surface in the cubic lattice to the 
diagonal hyperplane.
In this way, the rhombus tiling possesses a natural height representation.
As the average slope of this surface determines the average densities of the different 
tiles and vice versa, we can parametrize the ensemble alternatively by some slope variables.

Let us describe this more formally.
Our vector space is $\RR^3$ with the standard scalar product $<\cdot,\cdot>$.
The diagonal hyperplane in $\RR^3$ can be described by its normal vector 
$e^{\bot}=\frac{1}{\sqrt{3}}(1,1,1)^t$.
We call the space spanned by $e^{\bot}$ {\em perp(endicular) space} $V^{\bot}$
 and the complementary hyperplane {\em physical space} $V^{\|}$.
Let us denote the projections of vectors $x\in\RR^3$ to these subspaces by $x^{\bot}$ resp.
$x^{\|}$.
The {\em perp strain tensor} $E$ is a linear map
\be E:V^{\|}\to V^{\bot} .\ee
$E$ defines a hyperplane in $\RR^3$.
By definition, $E \equiv 0$ corresponds to $V^{\|}$.
We define matrix elements $E_1, E_2$ of $E$:
\be 
E_1 e_{1}^{\bot}=E e_{1}^{\|}, \qquad E_2 e_{2}^{\bot}=E e_{2}^{\|}. 
\ee
A normal vector to this hyperplane is given by $(1-E_1,1-E_2,1+E_1+E_2)^t$.

We can easily relate the average slope of a surface parametrized by $E=(E_1,E_2)$ to the mean
densities of the projected tiles:
\be 
\rho_1=\frac{1}{3}(1-E_1), \qquad \rho_2=\frac{1}{3}(1-E_2). 
\ee
The quadratic form  (\ref{form:rhombquad}) can
(with $r=\rho-\frac{1}{3}$) be rewritten in the form
\be 
s_2(E,E)= - \frac{1}{2}\cdot\frac{2}{\sqrt{3}}\cdot\frac{\pi}{9}\cdot 
(E_1^2+E_1E_2+E_2^2).
\ee
This is the expression given by Henley \cite{Hen88,Hen91} up to a constant of 
$\frac{2}{\sqrt{3}}$ which arises from different normalizations of unit vectors 
in physical and perp space in his setup.

\end{document}